\documentclass[article,aps,nofootinbib,twocolumn,superscriptaddress]{revtex4-1}
\input epsf
\usepackage{graphics}
\usepackage{amsmath}
\usepackage{amssymb}
\usepackage{bm}
\usepackage{color}
\usepackage{dcolumn}
\usepackage{hyphenat}
\usepackage{url}
\usepackage{comment}
\usepackage{float}
\usepackage[dvipsnames]{xcolor}

\allowdisplaybreaks

\def\be{\begin{equation}}
\def\ee{\end{equation}}
\def\ba{\begin{eqnarray}}
\def\ea{\end{eqnarray}}

\def\f{\frac}

\def\hub{{\mathcal H}}

\def\ie{{\frenchspacing\it i.e.~}}
\def\eg{{\frenchspacing\it e.g.~}}

\newcommand{\CITE}[1]{[{\bf CITE}]}

\usepackage{graphicx}

\begin{document}

\title{Phenomenology of Large Scale Structure in scalar-tensor theories: \\
joint prior covariance of $w_{\textrm{DE}}$, $\Sigma$ and $\mu$ in Horndeski}

\author{Juan Espejo} \affiliation{Leiden Observatory, Leiden University, Leiden 2300 RA, The Netherlands} \affiliation{Institute Lorentz, Leiden University, PO Box 9506, Leiden 2300 RA, The Netherlands}
\author{Simone Peirone} \affiliation{Institute Lorentz, Leiden University, PO Box 9506, Leiden 2300 RA, The Netherlands}
\author{Marco Raveri} \affiliation{Kavli Institute for Cosmological Physics, Enrico Fermi Institute,The University of Chicago, Chicago, Illinois 60637, USA}
\author{Kazuya Koyama}\affiliation{Institute of Cosmology and Gravitation, University of Portsmouth, Portsmouth, PO1 3FX, UK}
\author{Levon Pogosian} \affiliation{Department of Physics, Simon Fraser University, Burnaby, BC, V5A 1S6, Canada} \affiliation{Institute of Cosmology and Gravitation, University of Portsmouth, Portsmouth, PO1 3FX, UK}
\author{Alessandra Silvestri} \affiliation{Institute Lorentz, Leiden University, PO Box 9506, Leiden 2300 RA, The Netherlands}

\begin{abstract}
Ongoing and upcoming cosmological surveys will significantly improve our ability to probe the equation of state of dark energy, $w_{\rm DE}$, and the phenomenology of Large Scale Structure. They will allow us to constrain deviations from the $\Lambda$CDM predictions for the relations between the matter density contrast and the weak lensing and the Newtonian potential, described by the functions $\Sigma$ and $\mu$, respectively. In this work, we derive the theoretical prior for the joint covariance of $w_{\rm DE}$, $\Sigma$ and $\mu$, expected in general scalar-tensor theories with second order equations of motion (Horndeski gravity), focusing on their time-dependence at certain representative scales. We employ Monte-Carlo methods to generate large ensembles of statistically independent Horndeski models, focusing on those that are physically viable and in broad agreement with local tests of gravity, the observed cosmic expansion history and the measurement of the speed of gravitational waves from a binary neutron star merger. We identify several interesting features and trends in the distribution functions of $w_{\rm DE}$,  $\Sigma$ and $\mu$, as well as in their covariances; we confirm the high degree of correlation between $\Sigma$ and $\mu$ in scalar-tensor theories. The derived prior covariance matrices will allow us to reconstruct jointly $w_{\rm DE}$,  $\Sigma$ and $\mu$ in a non-parametric way. 
\end{abstract}

\maketitle

\section{Introduction}
\label{sec:introduction}
Among the primary goals of ongoing and future surveys of Large Scale Structure (LSS) is testing gravity on cosmological scales and shedding light on the nature of dark energy (DE), \ie the mysterious component thought to be sourcing cosmic acceleration~\cite{Riess:1998cb,Perlmutter:1998np,Silvestri:2009hh}. To this extent, they will provide accurate measurements of the effective equation of state of all non-dust contributions to the Friedmann equation at late times, hereafter referred to as $w_{\rm DE}(z)$. They will also measure deviations of the phenomenology of LSS from predictions of the standard model of cosmology, $\Lambda$CDM. These potential deviations are commonly encoded in the phenomenological functions $\Sigma$ and $\mu$ that parametrize  modifications of the perturbed Einstein's equations relating the matter density contrast to the lensing and the Newtonian potential, respectively~\cite{Amendola:2007rr,Bertschinger:2008zb,Pogosian:2010tj}. Stage IV LSS missions will provide constraints of order $1\%$ on $w_{\rm DE}(z)$, and of order $1-10\%$ on $\Sigma$ and $\mu$~\cite{Crittenden:2005wj,Zhao:2008bn,Zhao:2009fn,Hojjati:2011xd,Asaba:2013mxj}. 

Constraining functions of redshift and, possibly, scale with data, necessarily involves making assumptions about their properties. Such assumptions can be manifested in a choice of a specific parametric form, which, however, can limit the ability to capture nontrivial features and, more generally, is prone to biasing the outcome.  Alternatively, one can reconstruct these functions non-parametrically, \eg by binning them in redshift. As Principal Component Analysis (PCA) studies have shown~\cite{Crittenden:2005wj,Zhao:2009fn,Hojjati:2011xd,Asaba:2013mxj}, while the upcoming missions can constrain several eigenmodes of $w_{\rm DE}(z)$, $\Sigma(z)$ and $\mu(z)$, many more will remain unconstrained, with values in neighbouring bins effectively being degenerate. A partial lifting of the degeneracy, sufficient to aid the reconstruction, can be achieved by introducing correlations between bins in the form of prior covariances, that can be directly combined with the data covariance matrix~\cite{Crittenden:2011aa}. While different techniques can be employed to construct these correlation priors~\cite{Crittenden:2011aa,Zhao:2012aw,Raveri:2017qvt,Zhao:2017cud,Wang:2018fng,Casas:2017eob}, it is desirable for them to be theoretically informed. In a previous work~\cite{Raveri:2017qvt}, some of the authors have derived the theoretical prior covariance matrix for $w_{\rm DE}$ predicted by general scalar-tensor theories with second order equations of motion, \ie the Horndeski gravity~\cite{Horndeski:1974wa,Deffayet:2011gz,Kobayashi:2011nu}. Here we extend this work by creating joint theoretical covariance matrices for $w_{\rm DE}$ along with the phenomenological functions $\Sigma$ and $\mu$. In any specific theory of gravity, the expansion history and the evolution of perturbations follow from the same fundamental Lagrangian and are not independent of each other. Having a (weak) joint prior covariance between them will allow to constrain them jointly in a theoretically consistent way, while not biasing the outcome.

As in~\cite{Raveri:2017qvt}, we employ the unifying effective field theory (EFT) approach to DE and modified gravity (MG), and create large ensembles of statistically independent Horndeski models via Monte Carlo techniques. While we always require the speed of gravity to be equal to the speed of light today, as recently indicated by the gravitational wave measurement from a neutron star merger~\cite{Jana:2017ost}, we also separately consider the two sub-classes of Horndeski models: Generalized Brans-Dicke (GBD), \ie models with a standard form for the scalar kinetic term, and Horndeski models in which the speed of gravity is the same as that of light at all times. We also include constraints on the gravitational coupling, coming from Cosmic Microwave Background (CMB) and Big Bang Nucleosynthesis (BBN) bounds as well as laboratory tests. Furthermore, following~\cite{Peirone:2017ywi}, we also impose a weak Gaussian prior on the background expansion history, in order to be broadly consistent with existing cosmological distance measurements. Finally, we impose conditions for the physical viability of the sampled models, mainly avoiding ghost and gradient instabilities of the theory.

Our simulations allow us to derive several statistical properties of the distributions of $w_{\rm DE}$, $\Sigma$ and $\mu$, such as their mean values, and distribution functions,  in bins of time. Of more practical use (for non-parametric recsontructions), we also obtain their joint covariances and the functional forms of their correlation functions within each subclass of model. We study the dependence of the statistical ensemble on the imposed theoretical priors and mild obsevrational constraints. We also identify trends in the covariances associated to the different sub-classes of theories, while generally confirming the high degree of correlation between $\Sigma$ and $\mu$ in scalar-tensor theories. 

This work is organised as follows: in Section~\ref{sec:muSigma} we briefly introduce $\Sigma$ and $\mu$, the EFT of DE and MG as well as the classes of theories considered in this work. In Section~\ref{sec:methodology} we discuss the methodology adopted in order to build the samples and covariance matrices, in Section~\ref{sec:results} we present our results and, finally, in Section~\ref{sec:discussion} we discuss and summarise the main results.
\section{Evolution of Large Scale Structure in Horndeski theories} 
\label{sec:muSigma}
To study the dynamics of LSS, it is sufficient to focus on scalar perturbations around the flat Friedmann-Lemaitre-Robertson-Walker (FLRW) metric. In the conformal Newtonian gauge, the perturbed metric is given by
\be
ds^{2} = -a^{2}(\tau)[(1+2\Psi (\tau ,\vec{x}))d\tau ^{2}-(1-2\Phi (\tau ,\vec{x}))d\vec{x} ^{2}].
\ee
The evolution of the metric potentials $\Phi$ and $\Psi$ is coupled to that of matter fields through Einsteins' equations.
At late times, relevant for dark energy studies, when shear stresses from radiation and neutrinos are negligible, one can parametrize relations between the Fourier transforms of $\Phi$, $\Psi$ and the matter density contrast using the following equations~\cite{Amendola:2007rr,Bertschinger:2008zb,Pogosian:2010tj}
\ba
\label{eq:mu}
&&k^2\Psi = -4\pi G \mu(a,k)a^2 \rho \Delta \,,\\
\label{eq:Sigma}
&&k^2(\Phi+\Psi) = -8\pi G \Sigma(a,k) a^2\rho \Delta  \, ,
\ea
where $\rho$ is the background matter density and $\Delta=\delta +3aHv/k$ is the comoving density contrast. 
The functions $\Sigma$ and $\mu$ are equal to one in $\Lambda$CDM, but generally would be functions of time and the Fourier number $k$ in models beyond $\Lambda$CDM. 

When coupled to the Euler and the continuity equations for matter, eqs. (\ref{eq:mu}) and (\ref{eq:Sigma}) form a closed system that can be solved to obtain the phenomenology of LSS on linear scales~\cite{Pogosian:2010tj}. For example, one can use the publicly available Einstein-Boltzmann solver \texttt{MGCAMB}~\footnote{http://aliojjati.github.io/MGCAMB/}~\cite{Zhao:2008bn,Hojjati:2011ix} 
to compute the complete set of observables for a given choice of $\Sigma$ and $\mu$. 

Since the photon trajectories are affected by the Weyl potential, $\Phi+\Psi$, the function $\Sigma$ is particularly well probed by measurements of the weak lensing of distant galaxies and CMB, as well as measurements of galaxy number counts through the so-called magnification bias~\cite{Moessner:1997qs,Hojjati:2011xd,Asaba:2013mxj}. On the other hand, $\mu$ is best probed by galaxy clustering and redshift space distortions~\cite{Song:2010fg,Simpson:2012ra,Asaba:2013mxj}, since they are largely determined by the Newtonian potential $\Psi$.

\subsection{$w_{\textrm{DE}}$, $\Sigma$ and $\mu$ in Horndeski gravity}
\label{sec:EFT}
A broad class of theories that includes the majority of DE and MG models studied in the literature is that of scalar-tensor theories with second order equations of motion, known as Horndeski gravity~\cite{Horndeski:1974wa,Deffayet:2011gz,Kobayashi:2011nu}. Our aim is to identify general trends and correlations in the evolution of the background and linear perturbations that are common to broad ranges of models within the Horndeski gravity class. For this purpose, we can use the unifying effective field theory of DE (EFT) framework, in which the action for the background and perturbations is given by~\cite{Gubitosi:2012hu,Bloomfield:2012ff,Gleyzes:2013ooa,Bloomfield:2013efa,Piazza:2013coa}
\ba
\mathcal{S} =&& \int d^4x \sqrt{-g}  \bigg\{ \frac{m_0^2}{2} \Omega(\tau)R + \Lambda(\tau) - c(\tau)\,a^2\delta g^{00}  \nonumber \\ 
+&& \gamma_1(\tau)\frac{m_0^2H_0^2}{2} \left(a^2 \delta g^{00} \right)^2
 - \gamma_2(\tau)\frac{m_0^2H_0}{2} \, a^2\delta g^{00}\,\delta {K}{^\mu_\mu}  \nonumber \\
+&& \gamma_3(\tau) \frac{m_0^2}{2}\left[\left( \delta {K}{^\mu_\mu}\right)^2 - \delta {K}{^\mu_\nu}\,\delta {K}{^\nu_\mu}-\frac{ a^2}{2} \delta g^{00}\,\delta \mathcal{R}\right]+	\ldots \bigg\}\nonumber\\
+&& S_{m} [g_{\mu \nu}, \chi_m ],
\label{EFT_action}
\ea
where $m_0^{-2} = 8\pi G$, and $\delta g^{00}$, $\delta {K}{^\mu_\nu}$, $\delta K$ and $\delta R^{(3)}$ are, respectively, the perturbations of the time-time component of the metric, the extrinsic curvature and its trace, and the three dimensional spatial Ricci scalar on the constant-time hypersurfaces. The action (\ref{EFT_action}) is written in terms of the conformal time, $\tau$, and in the unitary gauge, in which the time coordinate is associated with hypersurfaces of a uniform scalar field. The functions $\Omega(\tau)$, $\Lambda(\tau)$ and $c(\tau)$ affect the evolution of the background and perturbations, with only two of them being independent as the third one can be derived using the Friedmann equations. The remaining functions, $\gamma_1$, $\gamma_2$ and $\gamma_3$, control the evolution of perturbations. Finally, $S_m$ is the action of all matter fields, $\chi_m$, that are minimally coupled to the metric $g_{\mu\nu}$. 

Given $\Omega(a)$ and $\Lambda(a)$, one can use the Friedmann equation to solve for the evolution of the Hubble parameter $\hub = a^{-1}da/d\tau$. Namely, introducing $y \equiv \hub^2$, we have 
\ba
\nonumber
\left(1+\Omega+\f{1}{2}a\Omega'\right)\f{d y}{d \ln a} &&+\left(1+\Omega+2a\Omega' +a^2\Omega'' \right)y \\
&&+\left( \frac{P_ma^2}{m_0^2} + \frac{\Lambda a^2}{m_0^2}\right) = 0\,,
\label{eq:hubble}
\ea
where the prime indicates differentiation with respect to the scale factor. Given the solution for $\hub(a)$, the effective DE EoS is defined via
\begin{align}
w_{\rm DE} \equiv \frac{P_{\rm DE}}{\rho_{\rm DE}} = \frac{-2\dot{\hub} -\hub^2 - P_m a^2/m_0^2}{ 3\hub^2 - \rho_m a^2/m_0^2 } \ ,
\label{eq:wDE}
\end{align}
where $\rho_m$ and $P_m$ are the combined energy density and the pressure of all particle species, and the over-dot denotes a derivative with respect to the conformal time. A more detailed description of the background solution is given in~\cite{Raveri:2017qvt}. 

To solve for the perturbations, in addition to $\Omega$ and $\Lambda$, one needs to specify  $\gamma_1$, $\gamma_2$ and $\gamma_3$ multiplying the second order terms in the action. From eq.~(\ref{EFT_action}), one can work out the full set of linearly perturbed Einstein equations for scalar, vector and tensor modes. Functions $\Omega$ and $\gamma_3$ affect both scalar and tensor perturbations. In particular, whenever $\gamma_3\neq 0$, the speed of gravity $c_T$ is different from the speed of light, $c=1$, making it a key phenomenological signature of Horndeski gravity.  It has become conventional to parameterize this difference as  $\alpha_T \equiv c_T^2-1$~\cite{Bellini:2014fua}, related to the EFT functions via
\be
\label{eq:alpha_t_equation}
\alpha_T=-\frac{\gamma_3}{1+\Omega+\gamma_3}  \, .
\ee
Such deviations have been severely constrained by the recent detection of the neutron star binary GW170817 and its electromagnetic counterpart GRB170817A~\cite{TheLIGOScientific:2017qsa,Monitor:2017mdv,Coulter:2017wya}, although one must keep two arguments in mind. Firstly, GW170817 is at a distance of $40$ Mpc, or $z \sim 0.01$, while cosmological data comes from higher redshifts. So, technically, one can have $\gamma_3\neq 0$ at $z>0.01$. Secondly, as pointed out in~\cite{deRham:2018red}, the GW170817 measurement was performed at energy scales that are close to the cut-off scale at which EFT actions, such as (\ref{EFT_action}), become invalid. As explicitly shown in~\cite{deRham:2018red} for Horndeski theories, one can have the speed of gravity differ from the speed of light at energy scales relevant for cosmology, but get restored to the speed of light at higher energies due to the terms that dominate near the cutoff scale.

Our aim is to study dynamics of linear scalar perturbations on sub-horizon scales, $k\gg aH$, targeted by Stage IV surveys such as Euclid. As shown in~\cite{Peirone:2017ywi}, to study the statistical properties of $\Sigma$ and $\mu$ on these scales, it is sufficient to work in the Quasi-Static Approximation (QSA), where the time-derivatives of the metric and the scalar field perturbations are neglected compared to their spatial gradients. Then, one can write an algebraic set of equations which can be solved to find the following analytical expressions for $\Sigma$ and $\mu$~\cite{Silvestri:2013ne,Pogosian:2016pwr}:
\be
\mu=\frac{m_{0}^{2}}{M_{*}^{2}}\frac{1+M^{2}a^{2}/k^{2}}{f_{3}/2f_{1}M_{*}^{2}+M^{2}(1+\alpha_{T})^{-1}a^{2}/k^{2}}
\label{eq:mu_qsa}
\ee
\be
\Sigma=\frac{m_{0}^{2}}{2M_{*}^{2}}\frac{1+f_{5}/f_{1}+M^{2}\left[1+(1+\alpha_{T})^{-1}\right]a^{2}/k^{2}}{f_{3}/2f_{1}M_{*}^{2}+M^{2}(1+\alpha_{T})^{-1}a^{2}/k^{2}},
\label{eq:sigma_qsa}
\ee
where $M$, $M_*$, $f_1$, $f_3$ and $f_5$ are given in the appendix. Note that, while $\gamma_1$ does not enter explicitly in the quasi-static expressions for $\Sigma$ and $\mu$, it still plays a role in determining the stability of perturbations~\cite{Kreisch:2017uet}.

\section{Methodology} 
\label{sec:methodology}
\begin{table}
\centering
\begin{tabular}{cl}
\textbf{Name$\qquad$}     & \textbf{EFT functions}    \\
\hline
GBD$\qquad$  & $\Omega$, $\Lambda$  \\
$H_S$ $\qquad$    & $\Omega$, $\Lambda$,  $\gamma_1$, $\gamma_2$   \\
Hor$\qquad$   & $\Omega$, $\Lambda$,  $\gamma_1$, $\gamma_2$,  $\gamma_3$ ($\gamma_3=0$ at $a=1$) \\
\hline
\end{tabular}
\caption{The classes of theories analyzed in this work along with the relevant EFT functions.}
\label{Classes_of_theories}
\end{table}

In our analysis, we will scan the theory space of Horndeski gravity by considering several representative combinations of EFT functions $\Omega(\tau)$, $\Lambda(\tau)$, $\gamma_1$, $\gamma_2$ and $\gamma_3$, with their time dependence drawn from a general ensemble.  Specifically, we will consider three families of scalar-tensor theories:
\begin{itemize}
\item Generalized Brans Dicke (GBD) models, \ie~theories with a standard  kinetic term for the scalar field.  Jordan Brans-Dicke~\cite{Brans:1961sx} and $f(R)$~\cite{Hu:2007nk} models are representatives of this class. Within the EFT framework, they require specifying two functions, $\Lambda$ and $\Omega$.
\item $H_S$: the subclass of theories in which the speed of gravity is the same as the speed of light. The $H_S$ class includes GBD models, and allows for non-canonical forms of the kinetic term for the scalar field but without the higher derivative couplings. Kinetic Gravity Braiding (KGB)~\cite{Deffayet:2010qz} is an example of such models. In the EFT language, it is described by four functions:  $\Lambda$,  $\Omega$, $\gamma_1$ and $\gamma_2$. We call this class of theories $H_S$ because it contains all Horndeski models in which the modifications with respect to $\Lambda$CDM are solely in the scalar (hence ``$S$'') sector (up to the modification of the friction term in the tensor equations from the non-minimal coupling).
\item Horndeski (Hor): refers to the entire class of scalar-tensor theories with second order equations of motion~\cite{Horndeski:1974wa}. It includes all terms in the action (\ref{EFT_action}) specified by functions: $\Lambda$,  $\Omega$, $\gamma_1$, $\gamma_2$ and $\gamma_3$. In Hor, the speed of gravity can be different from the speed of light, but we only allow such deviations at earlier epochs, requiring that $\gamma_3=0$ today to satisfy the constraint coming from the recent detection of gravitational waves from a neutron star binary along with the electromagnetic counterpart~\cite{Blas:2016qmn,Abbott:2016blz,Monitor:2017mdv}.
\end{itemize}
The above classes of theories, and the associated EFT functions, are summarized in Table~\ref{Classes_of_theories}.

In order to scan the theory space, we have adopted the numerical framework developed in~\cite{Raveri:2017qvt,Peirone:2017ywi}. It consists of a Monte Carlo (MC) code which samples the space of the EFT functions, building a statistically significant  ensemble of viable models. For each model, it computes and stores the values of $w_{\rm DE}$,  $\Sigma$ and $\mu$ at densely spaced values of redshifts.

To build the samples, we parametrize the EFT functions using a Pad\'e expansion of order $[M,N]$, \eg
\be
f(a)=\frac{\sum_{n=1}^{N}\alpha_{n}\left(a-a_{0}\right)^{n-1}}{1+\sum_{m=1}^{M}\beta_{m}\left(a-a_{0}\right)^{m}}  \,.
\ee 
We have progressively increased the truncation orders $M$ and $N$ and found that the results stabilize beyond $N=5$ and $M=4$. Hence, we set $M$ and $N$ at these values, giving $N+M=9$ free parameters for each EFT function. 

In our MC method, all the coefficients of each function change at each Monte Carlo step. This ensures that we get numerical results that go through the whole parameter space homogeneously.  We aim at having ensembles of $\sim 10^4$ viable models for each of the classes of theories discussed at the beginning of this section. We use expansions around $a_0=0$ and $a_0=1$, to represent models that are close to $\Lambda$CDM in the past ({\it thawing}) or at present ({\it freezing}), respectively. Since the acceptance rate is different in each case, the desired sample size is not reached after the same number of sampled models, leading to a different respective statistical significance. We address this by re-weighting the samples based on the respective acceptance rate when processing the data.

The MC sampler varies the coefficients $\alpha_{n}$ and $\beta_{m}$ in the range $[-1,1]$. We have investigated using broader ranges, such as $[-10,10]$ and $[-50,50]$, and found that it did not noticeably increase the ensemble of viable models. We attribute this to the fact that models with larger departures from LCDM are less likely to satisfy the stability constraints described below. We also vary the relevant cosmological parameters: the matter density fraction $\Omega_m \in [0,1]$, the DE density fraction $\Omega_{\rm DE} \in [0,1]$ and the Hubble parameter $H_0 \in [20,100]$ km/s/Mpc. 

To compute the background evolution for a given model, the sampler was interfaced with the Einstein-Boltzmann solver \texttt{EFTCAMB}~\cite{Hu:2013twa,Raveri:2014cka}, a publicly available patch to the \texttt{CAMB} code~\cite{Lewis:1999bs}. 
Given the background solution, the code applies the built-in \texttt{EFTCAMB} stability filters that check for ghost and gradient instabilities for the scalar and the tensor mode perturbations~\cite{2014arXiv1405.3590H}. In addition to that, in certain cases, we impose very weak observational and experimental priors on $\Omega$ and $H$, to exclude models that are in gross disagreement with known constraints, and require $\alpha_T=0$ today. These conditions are itemized as follows:
\begin{enumerate}
\item[C1.] $|\Omega(z^*) -1| <0.1$ at $z^*=1100$ and $z^*=0$, to be broadly consistent with the Big Bang Nucleosynthesis (BBN) and Cosmic Microwave Background (CMB) bounds~\cite{Uzan:2010pm} and the fifth-force constraints~\cite{Brax:2012gr,Joyce:2014kja,Perenon:2015sla};
\item[C2.] $\gamma_3 (z=0) =0$, which implies $\alpha_T(z=0) = 0$, consistent with the multi-messenger detection of the binary neutron star merger event GW170817 and GRB170817A~\cite{TheLIGOScientific:2017qsa,Monitor:2017mdv,Coulter:2017wya};
\item[C3.] a weak gaussian prior on $H(z)$ at redshifts corresponding to the existing angular diameter distance measurements from Baryon Acoustic Oscillations (BAO). This prior is built by setting the mean to the $H(z)$ reconstructed from Planck2015 best fit $\Lambda$CDM model~\cite{Ade:2015xua} and taking the standard deviation to be $30 \%$ of the mean value (see~\cite{Peirone:2017ywi} for further details);
\item[C4.] a very weak observational constraint coming from supernovae (SN) luminosity distance measurements~\cite{2012ApJ...746...85S}, with a significantly inflated covariance (by a factor of four) to avoid the biasing of our results by tensions between cosmological datasets. In order to impose this condition, we make use of the Monte Carlo Markov Chains sampler \texttt{COSMOMC}~\cite{Lewis:2002ah}. We do not impose this condition when computing the covariances of $w_{\rm DE}$, $\Sigma$ and $\mu$, since they are meant to be purely theoretical priors.
\end{enumerate}

After computing the cosmology, filtering out the models that are physically unviable, and imposing a given set of conditions C1-C4 (depending on the scenario under consideration), we compute the values of $w_{\textrm{DE}}(a)$, $\Sigma(a,k)$ and $\mu(a,k)$ using eqs.~(\ref{eq:wDE}), (\ref{eq:mu_qsa}) and (\ref{eq:sigma_qsa}). We test the sampler by computing $\Sigma$ and $\mu$ at fixed scales $k\in {\{0.01,0.085,0.15\}}$ $h$ Mpc$^{-1}$, not finding any significant dependence on $k$. This is because the sampled models tend to have a mass scale $M \sim H$, and thus the mass term has no effect on scales inside the horizon. In principle, one could also find models where $M \gg H$, but that would require a taylored sampling strategy, as this class of models is of measure zero in our framework~\cite{Peirone:2017ywi}. Models with $M \gg H$ include $f(R)$ and other chameleon type theories, which can be tested directly using simpler techniques. Our aim here is different: rather than constraining specific classes of theories, we want to derive weak priors that would allow us to directly reconstruct $w_{\textrm{DE}}$, $\Sigma$ and $\mu$ from the data. Such reconstructions will certainly allow for models with $M \gg H$.

We set $k=0.01$ $h$ Mpc$^{-1}$ for all the simulations, which ensures that linear theory holds well and, the QSA is valid~\cite{Peirone:2017ywi}. We store $w_{\textrm{DE}}(a)$, $\Sigma(a)$ and $\mu(a)$ at $100$ uniformly spaced values of $a \in [0.1,1]$, which corresponds to $z \in [0,9]$ and build ensembles of their values based on $\sim 10^4$ accepted models in each case. Given the ensembles, we compute the mean values and the covariance matrices of the $w_{\rm DE}(a)$, $\Sigma(a)$ and $\mu(a)$ bins. The covariance matrix is defined as
\be
\label{eq:covariance_matrix}
C_{ij}=\frac{1}{N_{\textrm{samp}}-1}\sum_{k =1}^{N_{\textrm{samp}}}\left(x_{i}^{(k)}-\bar{x}_{i}\right)\left(x_{j}^{(k)}-\bar{x}_{j}\right) \, ,
\ee
where $x^{(k)}_i=x^{(k)}(z_i)$,  $\bar{x}_{i}$ is the mean value of $x$ in the $i$-th redshift bin, and $k$ labels a member of the sample of $N_{\textrm{samp}}$ models in the ensemble. The prior covariance matrices, along with the mean values, can be used to build a Gaussian prior probability distribution function that can be used to reconstruct~\cite{Crittenden:2011aa} functions $w_{\rm DE}(a)$, $\Sigma(a)$ and $\mu(a)$ from data, as was done for $w_{\textrm{DE}}(a)$ in~\cite{Zhao:2012aw,Zhao:2017cud}. One can also define the normalized correlation matrix as
\be
\label{eq:correlation_matrix}
\mathcal{C}_{ij}=\frac{C_{ij}}{\sqrt{C_{ii}C_{jj}}}  \, .
\ee

For practical applications, it can be useful to have analytical expressions for the continuous correlation function defined as 
\be
\label{eq:correlation}
\mathcal{C}(a,a')\equiv \left\langle [x(a)-\bar{x}(a)][x(a)-\bar{x}(a')] \right\rangle.
\ee
We will derive them by fitting representative functional forms to (\ref{eq:correlation_matrix}).

\section{Results} 
\label{sec:results}

\begin{figure*}[!htbp]
\centering
 \includegraphics[width=13cm]{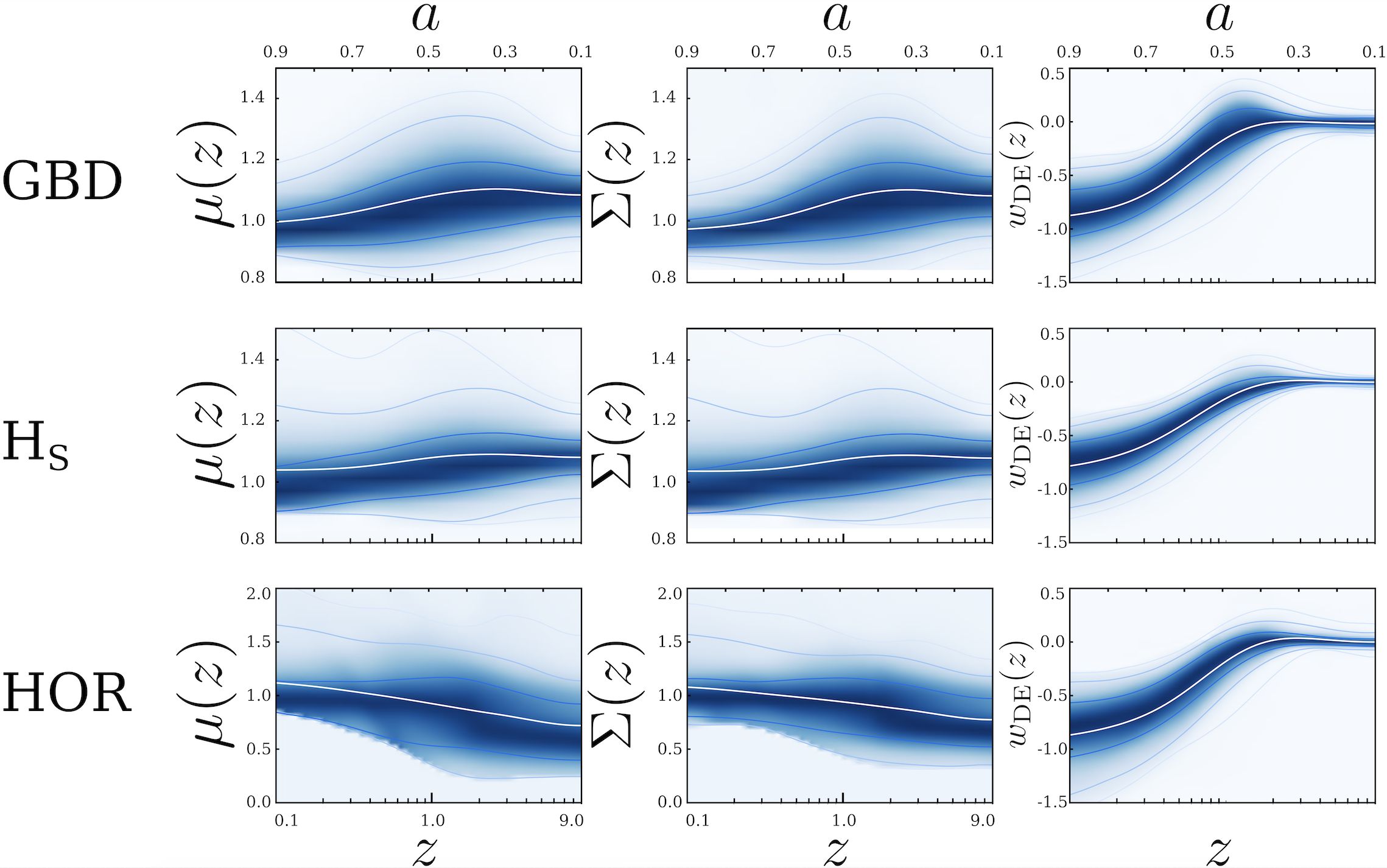}
\caption{The mean values (white line) and the 68\%, 95\% and 99\% confidence levels (solid blue lines) for $\mu(z)$, $\Sigma(z)$ and $w_{\textrm{DE}}(z)$ (left to right respectively) for the three classes of models: GBD (top row), $H_S$ (middle row) and Horndeski (bottom row). The shaded blue regions represent the probability distribution function (PDF) of each bin. These results are obtained from the simulations with conditions C1, C2, C3 and C4 from Sec.~\ref{sec:methodology} imposed.}
\label{fig:Means for all models}
\end{figure*}

The aim of this work is to provide theoretical prior distributions of bins of $w_{\textrm{DE}}(z)$, $\Sigma(z)$ and $\mu(z)$ that can be used in their joint reconstruction from data. Gaussian prior distributions can be build from the mean values and covariances of the bins obtained using the MC method described in the previous Section. As presented below, we have derived them separately for the three representative subclasses of Horndeski theories. The actual values of the means and the exact shapes of the prior distributions are not crucial in the Bayesian reconstruction method of~\cite{Crittenden:2011aa}. The role of the prior is to gently guide the reconstruction in regions of the parameter space poorly constrained by data.  For this reason, we also derive the approximate analytical forms describing the correlation between the bins that can be readily applied in practical applications without a loss of accuracy. 

\subsection{The mean values of the $w_{\textrm{DE}}$, $\Sigma$ and $\mu$ bins}

The mean values of $w_{\textrm{DE}}(z)$, $\Sigma(z)$ and $\mu(z)$ bins, along with the corresponding 68\%, 95\% and 99\% confidence level intervals, are shown in Fig.~\ref{fig:Means for all models}. They are obtained while imposing all four conditions, C1, C2, C3 and C4, of Sec.~\ref{sec:methodology}.

We observe that the mean values of $\Sigma$ and $\mu$ do not change significantly with redshift, and that for $H_S$ and Hor models they always remain within $\sim 1\sigma$ range of their $\Lambda$CDM values of $1$. For GBD, the LCDM values remain in the 2$\sigma$ range, with a clear trend towards values below $1$. This is because, in GBD, the values of $\Sigma$ and $\mu$ are largely determined by the prefactor $m_0^2/M_*^2=1/(1+\Omega)$ multiplying them both. Given a uniformly sampled $\Omega$, this prefactor is likely to be $<1$, because $1+\Omega$ must remain positive to guarantee the stability of the background solution, hence values of $\Omega \sim -1$ are often rejected by the stability filters built into the sampler. We note that, ultimately, the mean values should not play a significant role in practical applications. The uncertainties in the mean values are more relevant as they are linked to the covariances between $w_{\textrm{DE}}$, $\Sigma$ and $\mu$ bins. Nevertheless, one does need some values to put in the Gaussian prior, and it is interesting to see what one gets from the ensembles. 

In the case of $w_{\textrm{DE}}(z)$, the means are close to the $\Lambda$CDM value of $w_{\textrm{DE}}=-1$ at lower redshifts, where the SN data plays a role. At higher redshifts, $w_{\textrm{DE}}(z)$ tends to approach zero because of the tendency of the effective DE fluid to track the dominant density component~\cite{Raveri:2017qvt}. 

\subsection{The covariance of $w_{\textrm{DE}}$, $\Sigma$ and $\mu$}

\begin{figure}[!htbp]
\centering
\includegraphics[width=8.5cm]{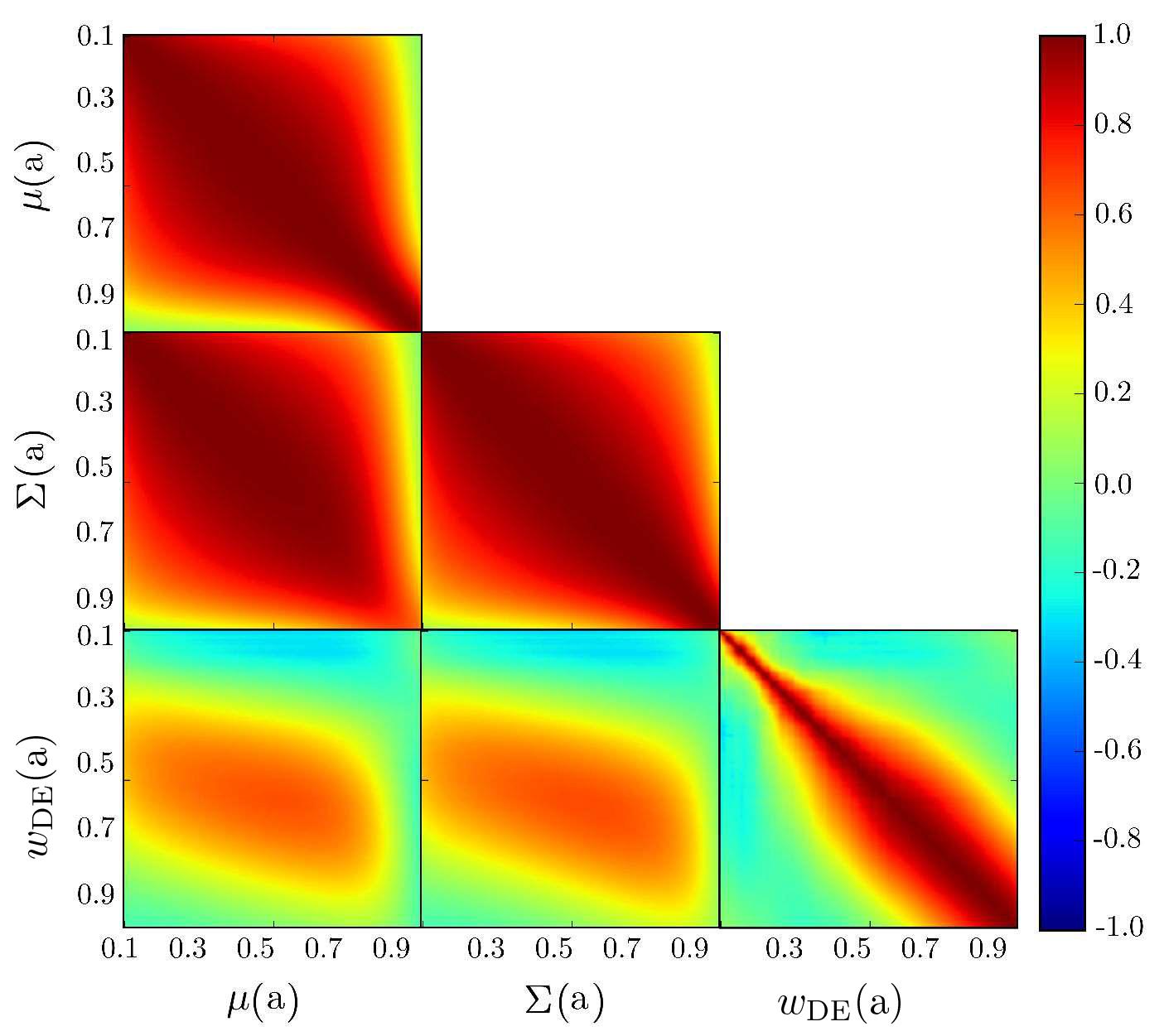}
\caption{ \label{fig:Correlations GBD}
Correlation matrices for the GBD class of models with conditions C1, C2 and C3 from Sec.~\ref{sec:methodology} imposed.}
\end{figure}
\begin{figure}[!htbp]
\centering
  \includegraphics[width=8.5cm]{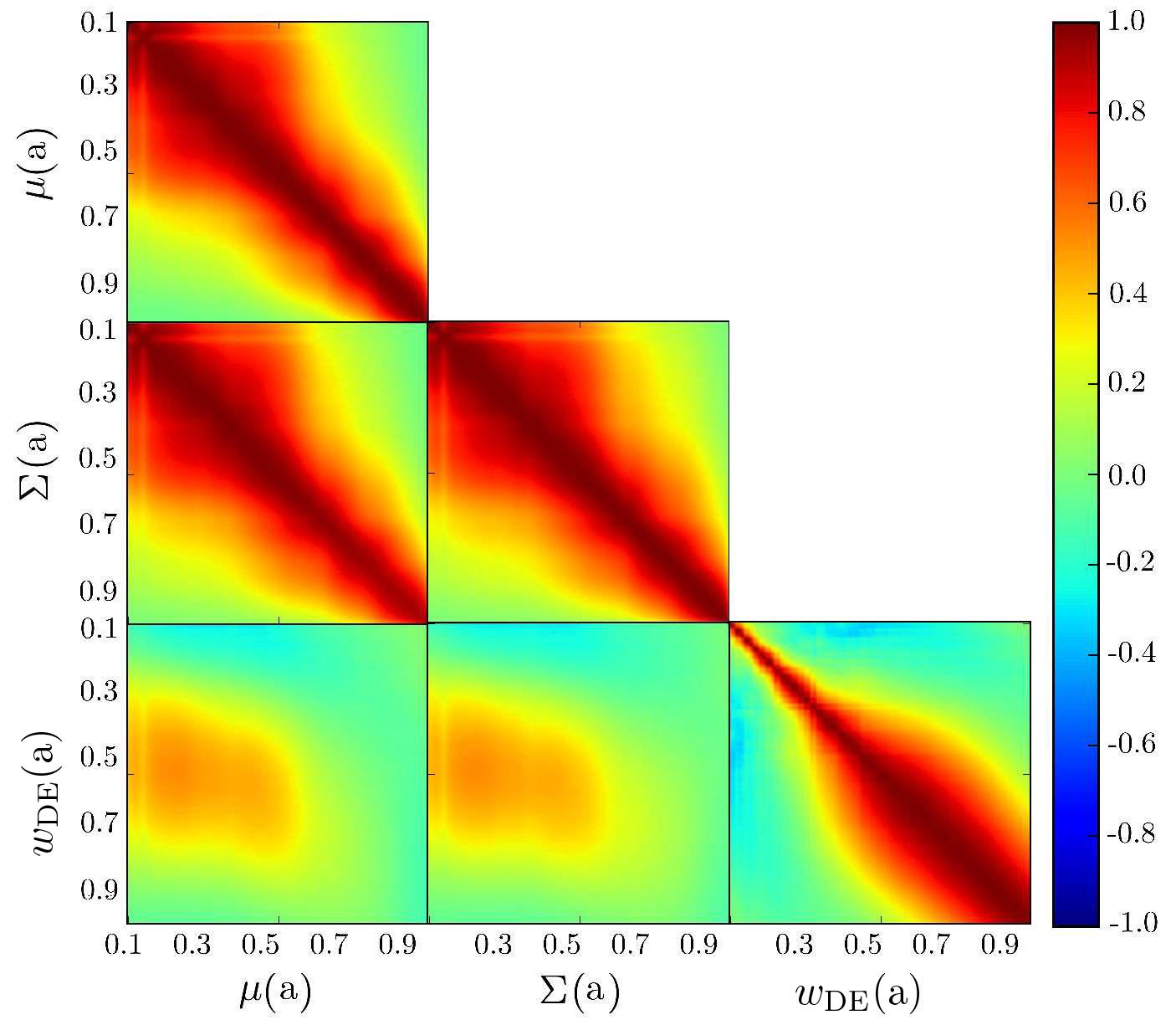}
\caption{  \label{fig:Correlations HS}
Correlation matrices for the $H_S$ class of models with conditions C1, C2 and C3 from Sec.~\ref{sec:methodology} imposed.}
\end{figure}
\begin{figure}[!htbp]
\centering
\includegraphics[width=8.5cm]{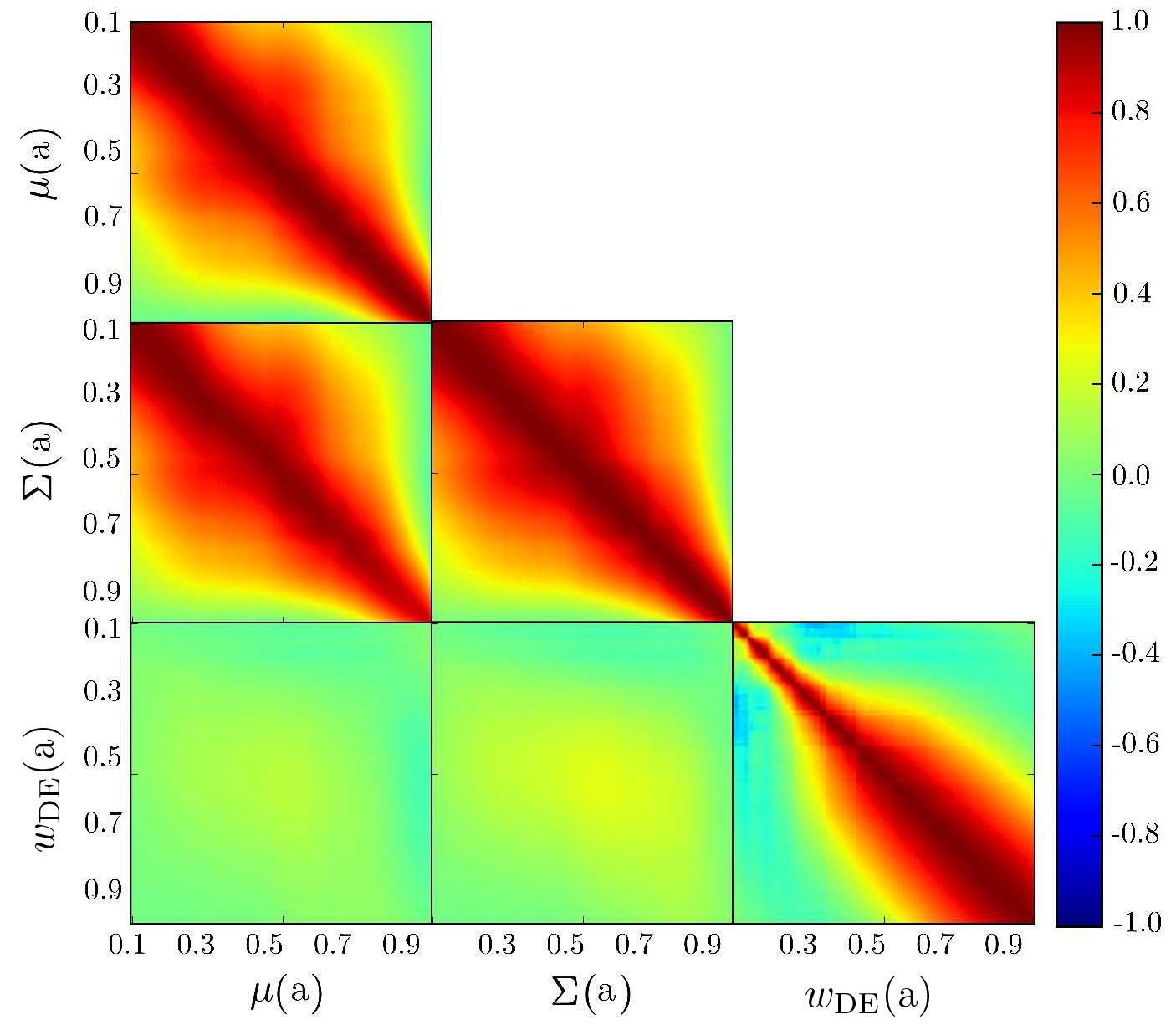}
\caption{
Correlation matrices for the Hor class of models with conditions C1, C2 and C3 from Sec.~\ref{sec:methodology} imposed.}
\label{fig:Correlations Hor}
\end{figure}

As mentioned earlier, the main goal of this work is to compute the covariance of the  $w_{\textrm{DE}}(z)$, $\Sigma(z)$ and $\mu(z)$ bins, so that it can be used as a theoretical prior in practical applications of the Bayesian reconstruction method~\cite{Crittenden:2011aa}. The covariances are computed using eq.~(\ref{eq:covariance_matrix}), while applying the conditions C1, C2, and C3 of Sec.~\ref{sec:methodology}. We do not include C4, as one should try not to use information from data in deriving the theoretical prior used in the Bayesian reconstruction. The covariances for each representative class of models are shown in Figs.~\ref{fig:Covariances GBD}, \ref{fig:Covariance HS} and \ref{fig:Covariances Hor} of the Appendix. While it is the covariances that are used in reconstructions, for the purpose of interpreting our results it is more informative to consider the correlation matrices computed using eq.~(\ref{eq:correlation_matrix}). They are shown in Figs.~\ref{fig:Correlations GBD}, \ref{fig:Correlations HS} and \ref{fig:Correlations Hor}  for the GBD, $H_S$ and Hor models, respectively. For each model, we display the correlations between the bins of the same function as well as the cross-correlations between different functions. 

Comparing Figs.~\ref{fig:Correlations GBD} and \ref{fig:Correlations HS}, one can clearly see that the correlation between $\Sigma$ and $\mu$ and between $\Sigma$/$\mu$ and $w_{\rm DE}$ is less pronounced in $H_S$, compared to the GBD case. This is due to the fact that more EFT functions participate in $H_S$. This trend continues only in part when one compares $H_S$ and Hor in Figs.~\ref{fig:Correlations HS} and \ref{fig:Correlations Hor}. Namely, the correlation $\Sigma$/$\mu$ and $w_{\rm DE}$ decreases, as expected, since Hor involves an additional EFT functions, $\gamma_3$. However, $\Sigma$ and $\mu$ are more correlated in Hor than they are in $H_S$. This is because $\gamma_3$ (equivalently $\alpha_T$) plays an important role in the stability constraints while being constrained by the condition C2, \ie $\gamma_3(z=0)=0$. The net effect of co-varying it with the other functions is to increase the level of correlation. To check this last point, we ran the same sampling imposing neither stability nor C2 and found that the correlation between  $\Sigma$ and $\mu$ decreases as expected when $\gamma_3$ is co-varied.

We note that, generally, the correlation between $\Sigma$ and $\mu$ is always significant, as also discussed in~\cite{Pogosian:2016pwr,Peirone:2017ywi}. This implies that, when constraining them within the framework of scalar-tensor theories, it does not make sense to fit these two functions to data independently. On the contrary, the cross correlation between $\Sigma$ and $\mu$ and $w_{\textrm{DE}}$ changes visibly for different classes of models. It is strong (up to $60$\%) in the GBD case: the two non-zero EFT functions, $\Omega$ and $\Lambda$, participate in the evolution of both the background and linear perturbations. For $H_S$ and Hor, in which the second order EFT functions $\gamma_i$ affect {\it only} the perturbations, this cross-correlation decreases. It is weak but still visible for H$_S$, and completely vanishes in for Hor. 

\subsection{Analytical forms of correlation functions}
\label{sec:correlations}

\begin{table*}
\centering
\begin{tabular}{c c c c}
		\multicolumn{4}{c}{\textbf{The best fit forms describing the correlations}}\\
		\hline
		& \textbf{$\mu$} & $\Sigma$ & $w_{\textrm{DE}}$\\
		\hline
		$\quad$&$\quad$&$\quad$&\\
		GBD & $\quad\left(1+\left(\left|\delta a\right|/0.65\right)^{2.25}\right)^{-1}$  & $\quad\left(1+\left(\left|\delta a\right|/0.7\right)^{2.2}\right)^{-1}$& $\quad\left(1+\left(\left|\delta \ln a\right|/0.29\right)^{3}\right)^{-1}$\\
		$H_S$ & $\quad\left(1+\left(\left|\delta a\right|/0.32\right)^{1.72}\right)^{-1}$  & $\quad\left(1+\left(\left|\delta a\right|/0.35\right)^{1.67}\right)^{-1}$& $\quad\left(1+\left(\left|\delta \ln a\right|/0.3\right)^{2.9}\right)^{-1}$\\
		Hor & $\quad\left(1+\left(\left|\delta a\right|/0.31\right)^{1.74}\right)^{-1}$  & $\quad\left(1+\left(\left|\delta a\right|/0.38\right)^{1.7}\right)^{-1}$& $\quad\left(1+\left(\left|\delta \ln a\right|/0.3\right)^{2.9}\right)^{-1}$\\
		\hline
\end{tabular}
\caption{The best fit analytical expressions of correlations of $\mu$, $\Sigma$ and $w_{\rm DE}$ for the three classes of models. For $\Sigma$ and $\mu$, the correlations depend on $|a-a^\prime|$, while for $w_{\rm DE}$ they scale with $|\ln a - \ln a^\prime|$, for the reasons explained in Sec.~\ref{sec:correlations}.}
	\label{Correlation length}
\end{table*}

In order to better interpret the numerically found correlation matrices, we fit them with simple analytical expressions.
In addition to providing insight into the time scaling of the correlations, they give a readily usable recipe for building correlation priors for practical applications~\cite{Zhao:2012aw,Zhao:2017cud,Wang:2018fng} when the numerically found covariances may not be available. 

Following the procedures of~\cite{Raveri:2017qvt} we use the generalized CPZ parametrization~\cite{Crittenden:2005wj}  given by
\be
\label{eq:CPZ parametrization}
\mathcal{C}(x,y)=\frac{1}{1+\left(\left|x-y\right|/\xi\right)^{n}} \,,
\ee
as well as other functional forms used in~\cite{Raveri:2017qvt} that, as we found, did not provide a better fit. We let the time coordinate, $x$ and $y$, be either the scale factor or $\ln{a}$. We select the best fit analytical form for the correlation by varying the exponent $n$ and the correlation length $\xi$ and minimizing the $\chi^2$. The CPZ form (\ref{eq:CPZ parametrization}), which happens to capture the features of our numerically found correlation matrices quite well, was designed to act as a low-pass Wiener filter. Namely, it assumes no prior correlation on widely separated time scales ($|x-y| > \xi$), allowing any slow, low-frequency, variations of the functions to pass through unbiased. On shorted time scales ($|x-y| < \xi$), however, any high-frequency variations will be suppressed, as the prior implies strong correlations between the neighbouring bins.

In the application of the correlation priors, the most important feature is their behaviour around the peak of the prior distribution. For this reason, for correlations of $\Sigma$ and $\mu$, we do not attempt to model the tails, and only fit the correlation $\mathcal{C}$ in the range $[x_p-\Delta x, x_p + \Delta x]$, where $x_p$ corresponds to the peak of the correlation at each value of $y$ and $\Delta x$ was chosen to be $\Delta a =0.2$. We do not fit the cross-correlations between different functions. In the case of $\Sigma$ and $\mu$, it is clear from Figs.~\ref{fig:Correlations GBD}, \ref{fig:Correlations HS} and \ref{fig:Correlations Hor} that their cross-correlations will have roughly the same functional form as the correlations. On the other hand, the cross-correlation between $w_{\rm DE}$ and $\Sigma$/$\mu$ is only relatively strong for GBD and can probably be ignored in practical applications when the numerically found covariances (shown in the Appendix \ref{app:covariance}) are not available.

The best fit functional forms of the correlations are shown in Table~\ref{Correlation length}. We notice that the time scaling for $\Sigma$ and $\mu$ correlations is in terms of $a$, while the  correlations for $w_{\textrm{DE}}$ scale with $\ln a$. This difference in scaling can be explained by observing that our sampling of the EFT functions is more or less uniform in $a$. The correlations of $\Sigma$ and $\mu$ retain the uniformity in $a$ because they directly depend on the EFT functions. In the case of $w_{\textrm{DE}}$, however, the non-minimal coupling of the scalar field leads to a tracking behaviour of the effective DE fluid, with its evolution dependent on the matter density $\rho_m \propto a^{-3}$. With the effective DE scaling as a power law of $a$, the correlations of its equation of state scale with $\ln a$. This scaling was also observed in~\cite{Raveri:2017qvt}, where it was also shown that for the minimally coupled scalar field, \ie the quintessence, the correlations scale as $a$, consistent with the above explanation.

In the case of $w_{\rm DE}$ correlations, the CPZ parameters $\xi$ and $n$ are approximately the same in the three classes of models, at $\xi \approx 0.3$ and $n \approx 3$. This is because the sampling of the background evolution depends mostly on the EFT functions $\Omega$ and $\Lambda$ in all three cases, and only indirectly depends on the $\gamma_i$ through the effect of stability conditions on model selection. 

For $\Sigma$ and $\mu$ correlations, there is a clear trend for correlations to become shorter range as one goes from GBD to $H_S$ and Hor. The correlation length is $\xi = 0.65$ for GBD, but $\xi \approx 0.3$ for $H_S$ and Hor. There is also a small change in the exponent from $n\approx 2.2$ to $n \approx 1.7$. The fact that the best fit forms of the correlation functions for H$_S$ and Hor are so similar suggests that the minor visible differences between Figs.~\ref{fig:Correlations HS} and~\ref{fig:Correlations Hor} concern mostly the tails of the correlation matrix, whereas our fits were performed near the peaks.

\section{Summary}
\label{sec:discussion}

We have derived joint theoretical priors for the effective DE equation of state $w_{\rm DE}$ and the phenomenological functions $\Sigma$ and $\mu$ within the Horndeski class of scalar-tensor theories, which includes all models with a single scalar field that have second order equations of motion in four dimensions. In order to do so, we worked within the unifying EFT framework and generated large ensembles of statistically independent models using Monte Carlo methods.

In our analysis, we separately considered the sub-class of GBD models, corresponding to theories with a standard kinetic term for the scalar field and a possible non-minimal coupling. We also considered the sub-class of Horndeski models with the speed of gravity equal to the speed of light at all times, which we dubbed $H_S$. Finally, we considered the class of Horndeski theories in which the speed of gravity is equal to the speed of light today, but not necessarily at higher redshifts, as indicated by the recent measurements of the GW from binary neutron star and its electromagnetic counterpart~\cite{TheLIGOScientific:2017qsa,Monitor:2017mdv,Coulter:2017wya}.  

Our priors are stored in the form of joint covariance matrices for binned $w_{\rm DE}$, $\Sigma$ and $\mu$. These matrices can be projected onto priors on parameters of any specific parametrization of these functions.

We spotted some notable differences in both the mean values and the covariances of $w_{\rm DE}$, $\Sigma$ and $\mu$ between the different classes of models depending on which constraints are imposed. For instance, we found that restrictions on the variation of the conformal coupling $\Omega$ (condition C1 of Sec.~\ref{sec:methodology}) directly impact the mean values of $\Sigma$ and $\mu$, and less directly the shape of $w_{\textrm{DE}}$. Furthermore, we found that C1, as well as the constraints on the speed of gravity (C2), have a bigger impact than the physical viability conditions built in \texttt{EFTCAMB}. 

We have identified simple analytical forms for the correlation functions, describing the correlations of $w_{\textrm{DE}}$, $\Sigma$ and $\mu$ at different redshifts, by fitting the CPZ parametrization (\ref{eq:CPZ parametrization}) to our numerical results. We noticed that in all the classes of models that we considered, the correlations of $\Sigma$ and $\mu$ scale with $|a-a^\prime|$, while for $w_{\rm DE}$ they scale with $|\ln a - \ln a^\prime|$. These analytical forms can be useful in practical applications of the Bayesian reconstruction method~\cite{Crittenden:2011aa}.

The prior covariances derived in this work can be used to perform a joint non-parametric reconstruction of $w_{\textrm{DE}}$, $\Sigma$ and $\mu$ from data similarly to the case done for $w_{\rm DE}$ in~\cite{Zhao:2012aw,Zhao:2017cud}. Introducing such joint correlation priors in the analysis will be essential in order to get significant constraints on the time evolution of the phenomenological functions within the context of scalar-tensor theories, while avoiding biasing the results by assuming specific functional forms. Such unbiased reconstructions would either constrain $\Lambda$CDM further, or perhaps point us towards an alternative theory of gravity.

\acknowledgements  AS and SP acknowledge support from the NWO and the Dutch Ministry of Education, Culture and Science (OCW), and also from the D-ITP consortium, a program of the NWO that is funded by the OCW. MR is supported by U.S.~Dept.~of Energy contract DE-FG02-13ER41958.  
The work of KK has received funding from the STFC grant ST/N000668/1, as well as the European Research Council (ERC) under the European Union's Horizon 2020 research and innovation programme (grant agreement 646702 ``CosTesGrav"). LP is supported by the Natural Sciences and Engineering Research Council (NSERC) of Canada. 
\appendix

\section{Supplementary equations}
\label{app:equations}

Under the QSA, functions defining the analytical expressions for $\Sigma$ and $\mu$ have the form~\citep{Bloomfield:2013efa}
\vspace{-0.5cm}

\ba
M^2 & = & C_{\pi} /f_1 \nonumber \\
M_*^2 &=& m_0^2\Omega + \bar{M}^2_2 \nonumber \\ 
f_1&=&C_3-C_1B_3 \nonumber \\
f_3&=&A_1(B_3C_2-B_1C_3)+A_2(B_1C_1-C_2) \nonumber \\
f_5&=&  B_3C_2-B_1C_3 \nonumber \\
\label{eq:f_i}
\ea
\vspace{-1cm}

where
\vspace{-0.5cm}

\ba
A_1 &=& 2(m_0^2\Omega+{\bar M}^2_2) \nonumber \\
A_2 &=& -m_0^2 \dot{\Omega}-\bar{M}^3_1 \nonumber \\
B_1 &=& -{ m_0^2\Omega +{\bar M}^2_2 \over m_0^2\Omega} \nonumber \\
B_3 &=& -{m_0^2 \dot{\Omega}  + (H + \partial_t){\bar M}^2_2  \over m_0^2 \Omega} \nonumber \\
C_1 &=& m_0^2 \dot{\Omega}  + (H + \partial_t){\bar M}^2_2 \nonumber \\
C_2 &=& - {1 \over 2}(m_0^2 \dot{\Omega}+\bar{M}^3_1)  \nonumber \\
C_3 &=& c-{1\over 2} (H+\partial_t)\bar{M}_1^3 + (H^2+\dot{H}+H\partial_t){\bar M}^2_2 \nonumber \\
C_\pi &=& {m_0^2\over 4}\dot{\Omega}\dot{R}^{(0)}-3c\dot{H} + {3\over 2} ( 3H\dot{H} +\dot{H} \partial_t+\ddot{H}) \bar{M}_1^3 
\nonumber \\
&+& 3 \dot{H}^2 \bar{M}_2^2 
\label{M2_Hornd}
\ea
and where we have used the dimensionful EFT functions and a dot indicates derivation w.r.t conformal time.

\section{Covariance matrices}
\label{app:covariance}

\begin{figure}[h]
\centering
\includegraphics[width=8.5cm]{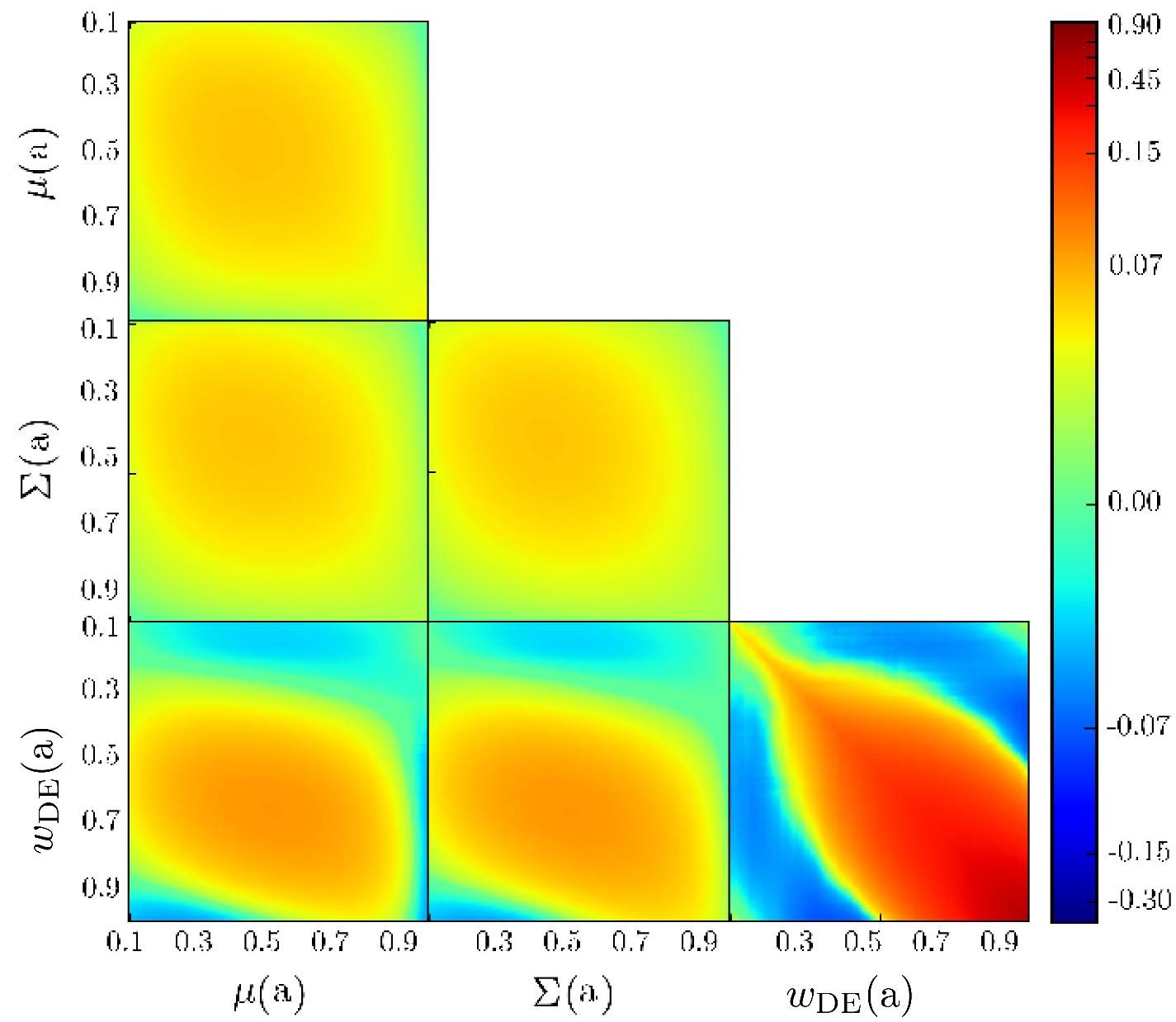}
\caption{\label{fig:Covariances GBD}
Covariance matrices for the GBD class of models with conditions C1, C2 and C3 from Sec.~\ref{sec:methodology} imposed.}
\end{figure}

\begin{figure}[h!]
\centering
\includegraphics[width=8.5cm]{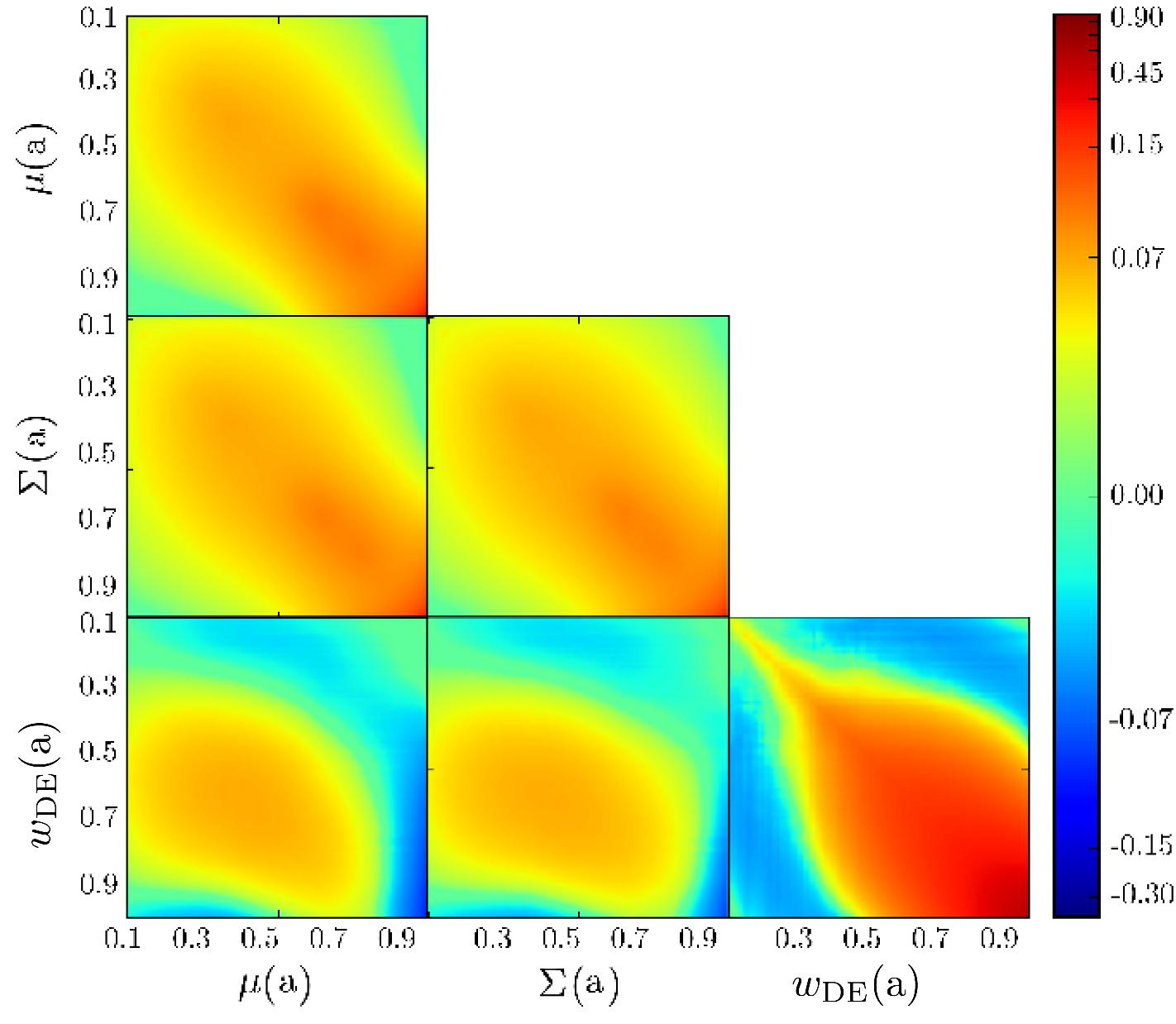}
\caption{\label{fig:Covariance HS}
Covariance matrices for the $H_S$ class of models with conditions C1, C2 and C3 from Sec.~\ref{sec:methodology} imposed.}
\end{figure}

\begin{figure}[b]
\centering
\includegraphics[width=8.5cm]{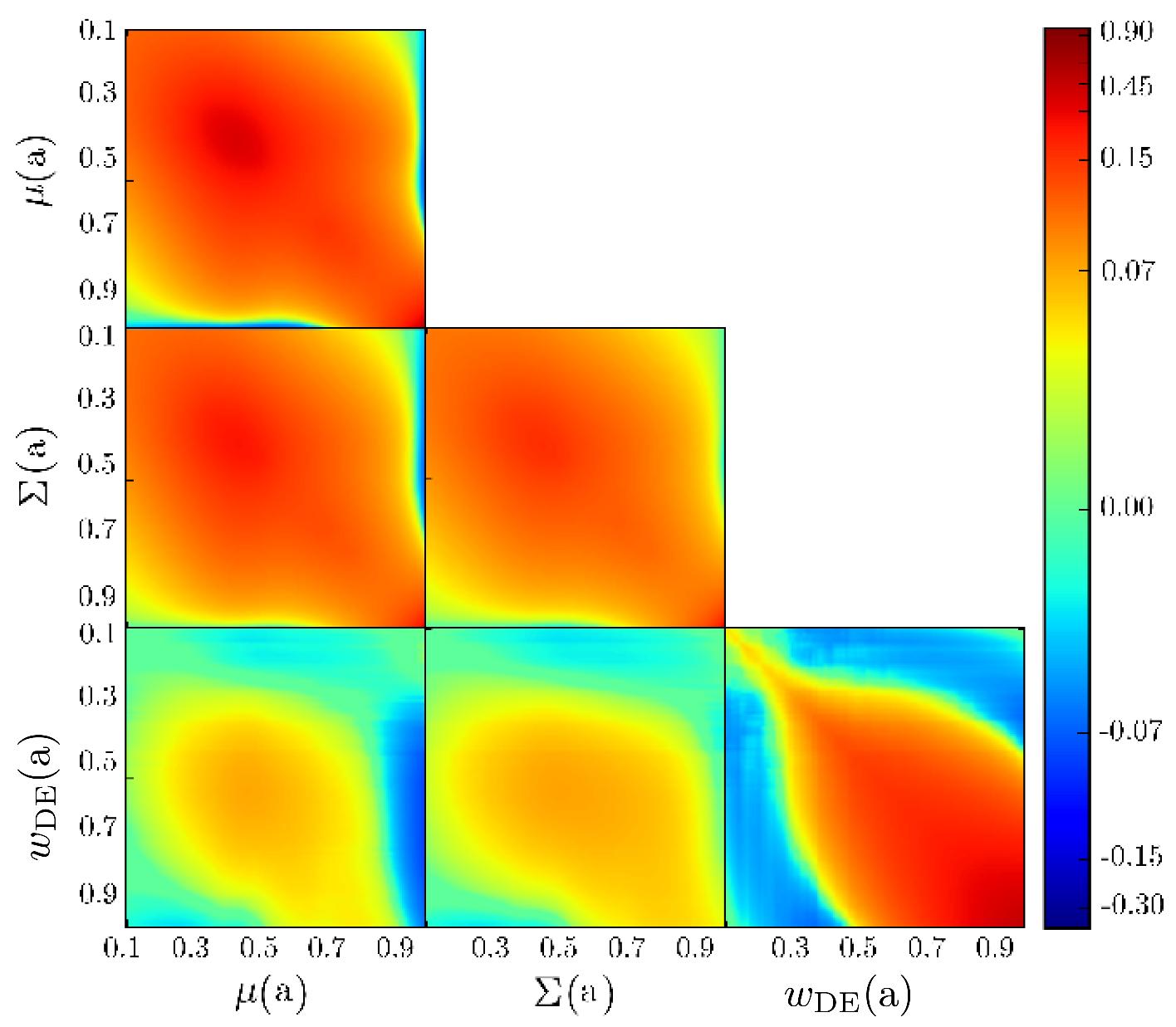}
\caption{  \label{fig:Covariances Hor}
Covariance matrices for the Hor class of models with conditions C1, C2 and C3 from Sec.~\ref{sec:methodology} imposed.}
\end{figure}

In practical applications of the Bayesian reconstruction method~\cite{Crittenden:2011aa}, one needs theoretical priors in the form of joint covariances of $w_{\textrm{DE}}(a)$, $\Sigma(a)$ and $\mu(a)$. These are shown in Figs.~\ref{fig:Covariances GBD}, \ref{fig:Covariance HS} and \ref{fig:Covariances Hor} for the three classes of models considered in this paper. They are obtained while applying the conditions C1, C2, and C3 of Sec.~\ref{sec:methodology}, and do not include C4, as the theoretical prior is not meant to be based on information from the data.

One can see that the prior variance in $w_{\textrm{DE}}(a)$ is smaller at higher redshifts and becomes larger towards $a=1$. This is because at higher redshifts, the effective DE tends to track the matter density, hence its equation of state is quite robustly close to zero. On the other hand, at lower redshifts, the effective DE fluid can develop its own independent dynamics as the matter density subsides, and there is more variation of possible $w_{\textrm{DE}}(a)$ histories within the ensemble.

The variances of $\Sigma$ and $\mu$ do not show a strong dependence on redshift, which is a reflection of the approximately uniform sampling of the EFT functions in $a$. The variances increase as one goes from GBD to $H_S$ to Hor, as expected, since the latter have a larger number of varied EFT functions that results in a larger scatter of $\Sigma$ and $\mu$ values.

\bibliography{EFTpriors}

\begin{thebibliography}{55}%
\makeatletter
\providecommand \@ifxundefined [1]{%
 \@ifx{#1\undefined}
}%
\providecommand \@ifnum [1]{%
 \ifnum #1\expandafter \@firstoftwo
 \else \expandafter \@secondoftwo
 \fi
}%
\providecommand \@ifx [1]{%
 \ifx #1\expandafter \@firstoftwo
 \else \expandafter \@secondoftwo
 \fi
}%
\providecommand \natexlab [1]{#1}%
\providecommand \enquote  [1]{``#1''}%
\providecommand \bibnamefont  [1]{#1}%
\providecommand \bibfnamefont [1]{#1}%
\providecommand \citenamefont [1]{#1}%
\providecommand \href@noop [0]{\@secondoftwo}%
\providecommand \href [0]{\begingroup \@sanitize@url \@href}%
\providecommand \@href[1]{\@@startlink{#1}\@@href}%
\providecommand \@@href[1]{\endgroup#1\@@endlink}%
\providecommand \@sanitize@url [0]{\catcode `\\12\catcode `\$12\catcode
  `\&12\catcode `\#12\catcode `\^12\catcode `\_12\catcode `\%12\relax}%
\providecommand \@@startlink[1]{}%
\providecommand \@@endlink[0]{}%
\providecommand \url  [0]{\begingroup\@sanitize@url \@url }%
\providecommand \@url [1]{\endgroup\@href {#1}{\urlprefix }}%
\providecommand \urlprefix  [0]{URL }%
\providecommand \Eprint [0]{\href }%
\providecommand \doibase [0]{http://dx.doi.org/}%
\providecommand \selectlanguage [0]{\@gobble}%
\providecommand \bibinfo  [0]{\@secondoftwo}%
\providecommand \bibfield  [0]{\@secondoftwo}%
\providecommand \translation [1]{[#1]}%
\providecommand \BibitemOpen [0]{}%
\providecommand \bibitemStop [0]{}%
\providecommand \bibitemNoStop [0]{.\EOS\space}%
\providecommand \EOS [0]{\spacefactor3000\relax}%
\providecommand \BibitemShut  [1]{\csname bibitem#1\endcsname}%
\let\auto@bib@innerbib\@empty
\bibitem [{\citenamefont {Riess}\ \emph {et~al.}(1998)\citenamefont {Riess}
  \emph {et~al.}}]{Riess:1998cb}%
  \BibitemOpen
  \bibfield  {author} {\bibinfo {author} {\bibfnamefont {A.~G.}\ \bibnamefont
  {Riess}} \emph {et~al.} (\bibinfo {collaboration} {Supernova Search Team}),\
  }\href {\doibase 10.1086/300499} {\bibfield  {journal} {\bibinfo  {journal}
  {Astron. J.}\ }\textbf {\bibinfo {volume} {116}},\ \bibinfo {pages} {1009}
  (\bibinfo {year} {1998})},\ \Eprint {http://arxiv.org/abs/astro-ph/9805201}
  {arXiv:astro-ph/9805201 [astro-ph]} \BibitemShut {NoStop}%
\bibitem [{\citenamefont {Perlmutter}\ \emph {et~al.}(1999)\citenamefont
  {Perlmutter} \emph {et~al.}}]{Perlmutter:1998np}%
  \BibitemOpen
  \bibfield  {author} {\bibinfo {author} {\bibfnamefont {S.}~\bibnamefont
  {Perlmutter}} \emph {et~al.} (\bibinfo {collaboration} {Supernova Cosmology
  Project}),\ }\href {\doibase 10.1086/307221} {\bibfield  {journal} {\bibinfo
  {journal} {Astrophys. J.}\ }\textbf {\bibinfo {volume} {517}},\ \bibinfo
  {pages} {565} (\bibinfo {year} {1999})},\ \Eprint
  {http://arxiv.org/abs/astro-ph/9812133} {arXiv:astro-ph/9812133 [astro-ph]}
  \BibitemShut {NoStop}%
\bibitem [{\citenamefont {Silvestri}\ and\ \citenamefont
  {Trodden}(2009)}]{Silvestri:2009hh}%
  \BibitemOpen
  \bibfield  {author} {\bibinfo {author} {\bibfnamefont {A.}~\bibnamefont
  {Silvestri}}\ and\ \bibinfo {author} {\bibfnamefont {M.}~\bibnamefont
  {Trodden}},\ }\href {\doibase 10.1088/0034-4885/72/9/096901} {\bibfield
  {journal} {\bibinfo  {journal} {Rept. Prog. Phys.}\ }\textbf {\bibinfo
  {volume} {72}},\ \bibinfo {pages} {096901} (\bibinfo {year} {2009})},\
  \Eprint {http://arxiv.org/abs/0904.0024} {arXiv:0904.0024 [astro-ph.CO]}
  \BibitemShut {NoStop}%
\bibitem [{\citenamefont {Amendola}\ \emph {et~al.}(2008)\citenamefont
  {Amendola}, \citenamefont {Kunz},\ and\ \citenamefont
  {Sapone}}]{Amendola:2007rr}%
  \BibitemOpen
  \bibfield  {author} {\bibinfo {author} {\bibfnamefont {L.}~\bibnamefont
  {Amendola}}, \bibinfo {author} {\bibfnamefont {M.}~\bibnamefont {Kunz}}, \
  and\ \bibinfo {author} {\bibfnamefont {D.}~\bibnamefont {Sapone}},\ }\href
  {\doibase 10.1088/1475-7516/2008/04/013} {\bibfield  {journal} {\bibinfo
  {journal} {JCAP}\ }\textbf {\bibinfo {volume} {0804}},\ \bibinfo {pages}
  {013} (\bibinfo {year} {2008})},\ \Eprint {http://arxiv.org/abs/0704.2421}
  {arXiv:0704.2421 [astro-ph]} \BibitemShut {NoStop}%
\bibitem [{\citenamefont {Bertschinger}\ and\ \citenamefont
  {Zukin}(2008)}]{Bertschinger:2008zb}%
  \BibitemOpen
  \bibfield  {author} {\bibinfo {author} {\bibfnamefont {E.}~\bibnamefont
  {Bertschinger}}\ and\ \bibinfo {author} {\bibfnamefont {P.}~\bibnamefont
  {Zukin}},\ }\href {\doibase 10.1103/PhysRevD.78.024015} {\bibfield  {journal}
  {\bibinfo  {journal} {Phys. Rev.}\ }\textbf {\bibinfo {volume} {D78}},\
  \bibinfo {pages} {024015} (\bibinfo {year} {2008})},\ \Eprint
  {http://arxiv.org/abs/0801.2431} {arXiv:0801.2431 [astro-ph]} \BibitemShut
  {NoStop}%
\bibitem [{\citenamefont {Pogosian}\ \emph {et~al.}(2010)\citenamefont
  {Pogosian}, \citenamefont {Silvestri}, \citenamefont {Koyama},\ and\
  \citenamefont {Zhao}}]{Pogosian:2010tj}%
  \BibitemOpen
  \bibfield  {author} {\bibinfo {author} {\bibfnamefont {L.}~\bibnamefont
  {Pogosian}}, \bibinfo {author} {\bibfnamefont {A.}~\bibnamefont {Silvestri}},
  \bibinfo {author} {\bibfnamefont {K.}~\bibnamefont {Koyama}}, \ and\ \bibinfo
  {author} {\bibfnamefont {G.-B.}\ \bibnamefont {Zhao}},\ }\href {\doibase
  10.1103/PhysRevD.81.104023} {\bibfield  {journal} {\bibinfo  {journal} {Phys.
  Rev.}\ }\textbf {\bibinfo {volume} {D81}},\ \bibinfo {pages} {104023}
  (\bibinfo {year} {2010})},\ \Eprint {http://arxiv.org/abs/1002.2382}
  {arXiv:1002.2382 [astro-ph.CO]} \BibitemShut {NoStop}%
\bibitem [{\citenamefont {Crittenden}\ \emph {et~al.}(2009)\citenamefont
  {Crittenden}, \citenamefont {Pogosian},\ and\ \citenamefont
  {Zhao}}]{Crittenden:2005wj}%
  \BibitemOpen
  \bibfield  {author} {\bibinfo {author} {\bibfnamefont {R.~G.}\ \bibnamefont
  {Crittenden}}, \bibinfo {author} {\bibfnamefont {L.}~\bibnamefont
  {Pogosian}}, \ and\ \bibinfo {author} {\bibfnamefont {G.-B.}\ \bibnamefont
  {Zhao}},\ }\href {\doibase 10.1088/1475-7516/2009/12/025} {\bibfield
  {journal} {\bibinfo  {journal} {JCAP}\ }\textbf {\bibinfo {volume} {0912}},\
  \bibinfo {pages} {025} (\bibinfo {year} {2009})},\ \Eprint
  {http://arxiv.org/abs/astro-ph/0510293} {arXiv:astro-ph/0510293 [astro-ph]}
  \BibitemShut {NoStop}%
\bibitem [{\citenamefont {Zhao}\ \emph
  {et~al.}(2009{\natexlab{a}})\citenamefont {Zhao}, \citenamefont {Pogosian},
  \citenamefont {Silvestri},\ and\ \citenamefont {Zylberberg}}]{Zhao:2008bn}%
  \BibitemOpen
  \bibfield  {author} {\bibinfo {author} {\bibfnamefont {G.-B.}\ \bibnamefont
  {Zhao}}, \bibinfo {author} {\bibfnamefont {L.}~\bibnamefont {Pogosian}},
  \bibinfo {author} {\bibfnamefont {A.}~\bibnamefont {Silvestri}}, \ and\
  \bibinfo {author} {\bibfnamefont {J.}~\bibnamefont {Zylberberg}},\ }\href
  {\doibase 10.1103/PhysRevD.79.083513} {\bibfield  {journal} {\bibinfo
  {journal} {Phys. Rev.}\ }\textbf {\bibinfo {volume} {D79}},\ \bibinfo {pages}
  {083513} (\bibinfo {year} {2009}{\natexlab{a}})},\ \Eprint
  {http://arxiv.org/abs/0809.3791} {arXiv:0809.3791 [astro-ph]} \BibitemShut
  {NoStop}%
\bibitem [{\citenamefont {Zhao}\ \emph
  {et~al.}(2009{\natexlab{b}})\citenamefont {Zhao}, \citenamefont {Pogosian},
  \citenamefont {Silvestri},\ and\ \citenamefont {Zylberberg}}]{Zhao:2009fn}%
  \BibitemOpen
  \bibfield  {author} {\bibinfo {author} {\bibfnamefont {G.-B.}\ \bibnamefont
  {Zhao}}, \bibinfo {author} {\bibfnamefont {L.}~\bibnamefont {Pogosian}},
  \bibinfo {author} {\bibfnamefont {A.}~\bibnamefont {Silvestri}}, \ and\
  \bibinfo {author} {\bibfnamefont {J.}~\bibnamefont {Zylberberg}},\ }\href
  {\doibase 10.1103/PhysRevLett.103.241301} {\bibfield  {journal} {\bibinfo
  {journal} {Phys. Rev. Lett.}\ }\textbf {\bibinfo {volume} {103}},\ \bibinfo
  {pages} {241301} (\bibinfo {year} {2009}{\natexlab{b}})},\ \Eprint
  {http://arxiv.org/abs/0905.1326} {arXiv:0905.1326 [astro-ph.CO]} \BibitemShut
  {NoStop}%
\bibitem [{\citenamefont {Hojjati}\ \emph {et~al.}(2012)\citenamefont
  {Hojjati}, \citenamefont {Zhao}, \citenamefont {Pogosian}, \citenamefont
  {Silvestri}, \citenamefont {Crittenden},\ and\ \citenamefont
  {Koyama}}]{Hojjati:2011xd}%
  \BibitemOpen
  \bibfield  {author} {\bibinfo {author} {\bibfnamefont {A.}~\bibnamefont
  {Hojjati}}, \bibinfo {author} {\bibfnamefont {G.-B.}\ \bibnamefont {Zhao}},
  \bibinfo {author} {\bibfnamefont {L.}~\bibnamefont {Pogosian}}, \bibinfo
  {author} {\bibfnamefont {A.}~\bibnamefont {Silvestri}}, \bibinfo {author}
  {\bibfnamefont {R.}~\bibnamefont {Crittenden}}, \ and\ \bibinfo {author}
  {\bibfnamefont {K.}~\bibnamefont {Koyama}},\ }\href {\doibase
  10.1103/PhysRevD.85.043508} {\bibfield  {journal} {\bibinfo  {journal} {Phys.
  Rev.}\ }\textbf {\bibinfo {volume} {D85}},\ \bibinfo {pages} {043508}
  (\bibinfo {year} {2012})},\ \Eprint {http://arxiv.org/abs/1111.3960}
  {arXiv:1111.3960 [astro-ph.CO]} \BibitemShut {NoStop}%
\bibitem [{\citenamefont {Asaba}\ \emph {et~al.}(2013)\citenamefont {Asaba},
  \citenamefont {Hikage}, \citenamefont {Koyama}, \citenamefont {Zhao},
  \citenamefont {Hojjati},\ and\ \citenamefont {Pogosian}}]{Asaba:2013mxj}%
  \BibitemOpen
  \bibfield  {author} {\bibinfo {author} {\bibfnamefont {S.}~\bibnamefont
  {Asaba}}, \bibinfo {author} {\bibfnamefont {C.}~\bibnamefont {Hikage}},
  \bibinfo {author} {\bibfnamefont {K.}~\bibnamefont {Koyama}}, \bibinfo
  {author} {\bibfnamefont {G.-B.}\ \bibnamefont {Zhao}}, \bibinfo {author}
  {\bibfnamefont {A.}~\bibnamefont {Hojjati}}, \ and\ \bibinfo {author}
  {\bibfnamefont {L.}~\bibnamefont {Pogosian}},\ }\href {\doibase
  10.1088/1475-7516/2013/08/029} {\bibfield  {journal} {\bibinfo  {journal}
  {JCAP}\ }\textbf {\bibinfo {volume} {1308}},\ \bibinfo {pages} {029}
  (\bibinfo {year} {2013})},\ \Eprint {http://arxiv.org/abs/1306.2546}
  {arXiv:1306.2546 [astro-ph.CO]} \BibitemShut {NoStop}%
\bibitem [{\citenamefont {Crittenden}\ \emph {et~al.}(2012)\citenamefont
  {Crittenden}, \citenamefont {Zhao}, \citenamefont {Pogosian}, \citenamefont
  {Samushia},\ and\ \citenamefont {Zhang}}]{Crittenden:2011aa}%
  \BibitemOpen
  \bibfield  {author} {\bibinfo {author} {\bibfnamefont {R.~G.}\ \bibnamefont
  {Crittenden}}, \bibinfo {author} {\bibfnamefont {G.-B.}\ \bibnamefont
  {Zhao}}, \bibinfo {author} {\bibfnamefont {L.}~\bibnamefont {Pogosian}},
  \bibinfo {author} {\bibfnamefont {L.}~\bibnamefont {Samushia}}, \ and\
  \bibinfo {author} {\bibfnamefont {X.}~\bibnamefont {Zhang}},\ }\href
  {\doibase 10.1088/1475-7516/2012/02/048} {\bibfield  {journal} {\bibinfo
  {journal} {JCAP}\ }\textbf {\bibinfo {volume} {1202}},\ \bibinfo {pages}
  {048} (\bibinfo {year} {2012})},\ \Eprint {http://arxiv.org/abs/1112.1693}
  {arXiv:1112.1693 [astro-ph.CO]} \BibitemShut {NoStop}%
\bibitem [{\citenamefont {Zhao}\ \emph {et~al.}(2012)\citenamefont {Zhao},
  \citenamefont {Crittenden}, \citenamefont {Pogosian},\ and\ \citenamefont
  {Zhang}}]{Zhao:2012aw}%
  \BibitemOpen
  \bibfield  {author} {\bibinfo {author} {\bibfnamefont {G.-B.}\ \bibnamefont
  {Zhao}}, \bibinfo {author} {\bibfnamefont {R.~G.}\ \bibnamefont
  {Crittenden}}, \bibinfo {author} {\bibfnamefont {L.}~\bibnamefont
  {Pogosian}}, \ and\ \bibinfo {author} {\bibfnamefont {X.}~\bibnamefont
  {Zhang}},\ }\href {\doibase 10.1103/PhysRevLett.109.171301} {\bibfield
  {journal} {\bibinfo  {journal} {Phys. Rev. Lett.}\ }\textbf {\bibinfo
  {volume} {109}},\ \bibinfo {pages} {171301} (\bibinfo {year} {2012})},\
  \Eprint {http://arxiv.org/abs/1207.3804} {arXiv:1207.3804 [astro-ph.CO]}
  \BibitemShut {NoStop}%
\bibitem [{\citenamefont {Raveri}\ \emph {et~al.}(2017)\citenamefont {Raveri},
  \citenamefont {Bull}, \citenamefont {Silvestri},\ and\ \citenamefont
  {Pogosian}}]{Raveri:2017qvt}%
  \BibitemOpen
  \bibfield  {author} {\bibinfo {author} {\bibfnamefont {M.}~\bibnamefont
  {Raveri}}, \bibinfo {author} {\bibfnamefont {P.}~\bibnamefont {Bull}},
  \bibinfo {author} {\bibfnamefont {A.}~\bibnamefont {Silvestri}}, \ and\
  \bibinfo {author} {\bibfnamefont {L.}~\bibnamefont {Pogosian}},\ }\href
  {\doibase 10.1103/PhysRevD.96.083509} {\bibfield  {journal} {\bibinfo
  {journal} {Phys. Rev.}\ }\textbf {\bibinfo {volume} {D96}},\ \bibinfo {pages}
  {083509} (\bibinfo {year} {2017})},\ \Eprint
  {http://arxiv.org/abs/1703.05297} {arXiv:1703.05297 [astro-ph.CO]}
  \BibitemShut {NoStop}%
\bibitem [{\citenamefont {Zhao}\ \emph {et~al.}(2017)\citenamefont {Zhao} \emph
  {et~al.}}]{Zhao:2017cud}%
  \BibitemOpen
  \bibfield  {author} {\bibinfo {author} {\bibfnamefont {G.-B.}\ \bibnamefont
  {Zhao}} \emph {et~al.},\ }\href {\doibase 10.1038/s41550-017-0216-z}
  {\bibfield  {journal} {\bibinfo  {journal} {Nat. Astron.}\ }\textbf {\bibinfo
  {volume} {1}},\ \bibinfo {pages} {627} (\bibinfo {year} {2017})},\ \Eprint
  {http://arxiv.org/abs/1701.08165} {arXiv:1701.08165 [astro-ph.CO]}
  \BibitemShut {NoStop}%
\bibitem [{\citenamefont {Wang}\ \emph {et~al.}(2018)\citenamefont {Wang},
  \citenamefont {Pogosian}, \citenamefont {Zhao},\ and\ \citenamefont
  {Zucca}}]{Wang:2018fng}%
  \BibitemOpen
  \bibfield  {author} {\bibinfo {author} {\bibfnamefont {Y.}~\bibnamefont
  {Wang}}, \bibinfo {author} {\bibfnamefont {L.}~\bibnamefont {Pogosian}},
  \bibinfo {author} {\bibfnamefont {G.-B.}\ \bibnamefont {Zhao}}, \ and\
  \bibinfo {author} {\bibfnamefont {A.}~\bibnamefont {Zucca}},\ }\href@noop {}
  {\  (\bibinfo {year} {2018})},\ \Eprint {http://arxiv.org/abs/1807.03772}
  {arXiv:1807.03772 [astro-ph.CO]} \BibitemShut {NoStop}%
\bibitem [{\citenamefont {Casas}\ \emph {et~al.}(2017)\citenamefont {Casas},
  \citenamefont {Kunz}, \citenamefont {Martinelli},\ and\ \citenamefont
  {Pettorino}}]{Casas:2017eob}%
  \BibitemOpen
  \bibfield  {author} {\bibinfo {author} {\bibfnamefont {S.}~\bibnamefont
  {Casas}}, \bibinfo {author} {\bibfnamefont {M.}~\bibnamefont {Kunz}},
  \bibinfo {author} {\bibfnamefont {M.}~\bibnamefont {Martinelli}}, \ and\
  \bibinfo {author} {\bibfnamefont {V.}~\bibnamefont {Pettorino}},\ }\href
  {\doibase 10.1016/j.dark.2017.09.009} {\bibfield  {journal} {\bibinfo
  {journal} {Phys. Dark Univ.}\ }\textbf {\bibinfo {volume} {18}},\ \bibinfo
  {pages} {73} (\bibinfo {year} {2017})},\ \Eprint
  {http://arxiv.org/abs/1703.01271} {arXiv:1703.01271 [astro-ph.CO]}
  \BibitemShut {NoStop}%
\bibitem [{\citenamefont {Horndeski}(1974)}]{Horndeski:1974wa}%
  \BibitemOpen
  \bibfield  {author} {\bibinfo {author} {\bibfnamefont {G.~W.}\ \bibnamefont
  {Horndeski}},\ }\href {\doibase 10.1007/BF01807638} {\bibfield  {journal}
  {\bibinfo  {journal} {Int. J. Theor. Phys.}\ }\textbf {\bibinfo {volume}
  {10}},\ \bibinfo {pages} {363} (\bibinfo {year} {1974})}\BibitemShut
  {NoStop}%
\bibitem [{\citenamefont {Deffayet}\ \emph {et~al.}(2011)\citenamefont
  {Deffayet}, \citenamefont {Gao}, \citenamefont {Steer},\ and\ \citenamefont
  {Zahariade}}]{Deffayet:2011gz}%
  \BibitemOpen
  \bibfield  {author} {\bibinfo {author} {\bibfnamefont {C.}~\bibnamefont
  {Deffayet}}, \bibinfo {author} {\bibfnamefont {X.}~\bibnamefont {Gao}},
  \bibinfo {author} {\bibfnamefont {D.~A.}\ \bibnamefont {Steer}}, \ and\
  \bibinfo {author} {\bibfnamefont {G.}~\bibnamefont {Zahariade}},\ }\href
  {\doibase 10.1103/PhysRevD.84.064039} {\bibfield  {journal} {\bibinfo
  {journal} {Phys. Rev.}\ }\textbf {\bibinfo {volume} {D84}},\ \bibinfo {pages}
  {064039} (\bibinfo {year} {2011})},\ \Eprint {http://arxiv.org/abs/1103.3260}
  {arXiv:1103.3260 [hep-th]} \BibitemShut {NoStop}%
\bibitem [{\citenamefont {Kobayashi}\ \emph {et~al.}(2011)\citenamefont
  {Kobayashi}, \citenamefont {Yamaguchi},\ and\ \citenamefont
  {Yokoyama}}]{Kobayashi:2011nu}%
  \BibitemOpen
  \bibfield  {author} {\bibinfo {author} {\bibfnamefont {T.}~\bibnamefont
  {Kobayashi}}, \bibinfo {author} {\bibfnamefont {M.}~\bibnamefont
  {Yamaguchi}}, \ and\ \bibinfo {author} {\bibfnamefont {J.}~\bibnamefont
  {Yokoyama}},\ }\href {\doibase 10.1143/PTP.126.511} {\bibfield  {journal}
  {\bibinfo  {journal} {Prog. Theor. Phys.}\ }\textbf {\bibinfo {volume}
  {126}},\ \bibinfo {pages} {511} (\bibinfo {year} {2011})},\ \Eprint
  {http://arxiv.org/abs/1105.5723} {arXiv:1105.5723 [hep-th]} \BibitemShut
  {NoStop}%
\bibitem [{\citenamefont {Jana}\ \emph {et~al.}(2017)\citenamefont {Jana},
  \citenamefont {Chakravarty},\ and\ \citenamefont {Mohanty}}]{Jana:2017ost}%
  \BibitemOpen
  \bibfield  {author} {\bibinfo {author} {\bibfnamefont {S.}~\bibnamefont
  {Jana}}, \bibinfo {author} {\bibfnamefont {G.~K.}\ \bibnamefont
  {Chakravarty}}, \ and\ \bibinfo {author} {\bibfnamefont {S.}~\bibnamefont
  {Mohanty}},\ }\href@noop {} {\  (\bibinfo {year} {2017})},\ \Eprint
  {http://arxiv.org/abs/1711.04137} {arXiv:1711.04137 [gr-qc]} \BibitemShut
  {NoStop}%
\bibitem [{\citenamefont {Peirone}\ \emph {et~al.}(2018)\citenamefont
  {Peirone}, \citenamefont {Koyama}, \citenamefont {Pogosian}, \citenamefont
  {Raveri},\ and\ \citenamefont {Silvestri}}]{Peirone:2017ywi}%
  \BibitemOpen
  \bibfield  {author} {\bibinfo {author} {\bibfnamefont {S.}~\bibnamefont
  {Peirone}}, \bibinfo {author} {\bibfnamefont {K.}~\bibnamefont {Koyama}},
  \bibinfo {author} {\bibfnamefont {L.}~\bibnamefont {Pogosian}}, \bibinfo
  {author} {\bibfnamefont {M.}~\bibnamefont {Raveri}}, \ and\ \bibinfo {author}
  {\bibfnamefont {A.}~\bibnamefont {Silvestri}},\ }\href {\doibase
  10.1103/PhysRevD.97.043519} {\bibfield  {journal} {\bibinfo  {journal} {Phys.
  Rev.}\ }\textbf {\bibinfo {volume} {D97}},\ \bibinfo {pages} {043519}
  (\bibinfo {year} {2018})},\ \Eprint {http://arxiv.org/abs/1712.00444}
  {arXiv:1712.00444 [astro-ph.CO]} \BibitemShut {NoStop}%
\bibitem [{\citenamefont {Hojjati}\ \emph {et~al.}(2011)\citenamefont
  {Hojjati}, \citenamefont {Pogosian},\ and\ \citenamefont
  {Zhao}}]{Hojjati:2011ix}%
  \BibitemOpen
  \bibfield  {author} {\bibinfo {author} {\bibfnamefont {A.}~\bibnamefont
  {Hojjati}}, \bibinfo {author} {\bibfnamefont {L.}~\bibnamefont {Pogosian}}, \
  and\ \bibinfo {author} {\bibfnamefont {G.-B.}\ \bibnamefont {Zhao}},\ }\href
  {\doibase 10.1088/1475-7516/2011/08/005} {\bibfield  {journal} {\bibinfo
  {journal} {JCAP}\ }\textbf {\bibinfo {volume} {1108}},\ \bibinfo {pages}
  {005} (\bibinfo {year} {2011})},\ \Eprint {http://arxiv.org/abs/1106.4543}
  {arXiv:1106.4543 [astro-ph.CO]} \BibitemShut {NoStop}%
\bibitem [{\citenamefont {Moessner}\ \emph {et~al.}(1998)\citenamefont
  {Moessner}, \citenamefont {Jain},\ and\ \citenamefont
  {Villumsen}}]{Moessner:1997qs}%
  \BibitemOpen
  \bibfield  {author} {\bibinfo {author} {\bibfnamefont {R.}~\bibnamefont
  {Moessner}}, \bibinfo {author} {\bibfnamefont {B.}~\bibnamefont {Jain}}, \
  and\ \bibinfo {author} {\bibfnamefont {J.~V.}\ \bibnamefont {Villumsen}},\
  }\href {\doibase 10.1046/j.1365-8711.1998.01225.x} {\bibfield  {journal}
  {\bibinfo  {journal} {Mon. Not. Roy. Astron. Soc.}\ }\textbf {\bibinfo
  {volume} {294}},\ \bibinfo {pages} {291} (\bibinfo {year} {1998})},\ \Eprint
  {http://arxiv.org/abs/astro-ph/9708271} {arXiv:astro-ph/9708271 [astro-ph]}
  \BibitemShut {NoStop}%
\bibitem [{\citenamefont {Song}\ \emph {et~al.}(2011)\citenamefont {Song},
  \citenamefont {Zhao}, \citenamefont {Bacon}, \citenamefont {Koyama},
  \citenamefont {Nichol},\ and\ \citenamefont {Pogosian}}]{Song:2010fg}%
  \BibitemOpen
  \bibfield  {author} {\bibinfo {author} {\bibfnamefont {Y.-S.}\ \bibnamefont
  {Song}}, \bibinfo {author} {\bibfnamefont {G.-B.}\ \bibnamefont {Zhao}},
  \bibinfo {author} {\bibfnamefont {D.}~\bibnamefont {Bacon}}, \bibinfo
  {author} {\bibfnamefont {K.}~\bibnamefont {Koyama}}, \bibinfo {author}
  {\bibfnamefont {R.~C.}\ \bibnamefont {Nichol}}, \ and\ \bibinfo {author}
  {\bibfnamefont {L.}~\bibnamefont {Pogosian}},\ }\href {\doibase
  10.1103/PhysRevD.84.083523} {\bibfield  {journal} {\bibinfo  {journal} {Phys.
  Rev.}\ }\textbf {\bibinfo {volume} {D84}},\ \bibinfo {pages} {083523}
  (\bibinfo {year} {2011})},\ \Eprint {http://arxiv.org/abs/1011.2106}
  {arXiv:1011.2106 [astro-ph.CO]} \BibitemShut {NoStop}%
\bibitem [{\citenamefont {Simpson}\ \emph {et~al.}(2013)\citenamefont {Simpson}
  \emph {et~al.}}]{Simpson:2012ra}%
  \BibitemOpen
  \bibfield  {author} {\bibinfo {author} {\bibfnamefont {F.}~\bibnamefont
  {Simpson}} \emph {et~al.},\ }\href {\doibase 10.1093/mnras/sts493} {\bibfield
   {journal} {\bibinfo  {journal} {Mon. Not. Roy. Astron. Soc.}\ }\textbf
  {\bibinfo {volume} {429}},\ \bibinfo {pages} {2249} (\bibinfo {year}
  {2013})},\ \Eprint {http://arxiv.org/abs/1212.3339} {arXiv:1212.3339
  [astro-ph.CO]} \BibitemShut {NoStop}%
\bibitem [{\citenamefont {Gubitosi}\ \emph {et~al.}(2013)\citenamefont
  {Gubitosi}, \citenamefont {Piazza},\ and\ \citenamefont
  {Vernizzi}}]{Gubitosi:2012hu}%
  \BibitemOpen
  \bibfield  {author} {\bibinfo {author} {\bibfnamefont {G.}~\bibnamefont
  {Gubitosi}}, \bibinfo {author} {\bibfnamefont {F.}~\bibnamefont {Piazza}}, \
  and\ \bibinfo {author} {\bibfnamefont {F.}~\bibnamefont {Vernizzi}},\ }\href
  {\doibase 10.1088/1475-7516/2013/02/032} {\bibfield  {journal} {\bibinfo
  {journal} {JCAP}\ }\textbf {\bibinfo {volume} {1302}},\ \bibinfo {pages}
  {032} (\bibinfo {year} {2013})},\ \bibinfo {note} {[JCAP1302,032(2013)]},\
  \Eprint {http://arxiv.org/abs/1210.0201} {arXiv:1210.0201 [hep-th]}
  \BibitemShut {NoStop}%
\bibitem [{\citenamefont {Bloomfield}\ \emph {et~al.}(2013)\citenamefont
  {Bloomfield}, \citenamefont {Flanagan}, \citenamefont {Park},\ and\
  \citenamefont {Watson}}]{Bloomfield:2012ff}%
  \BibitemOpen
  \bibfield  {author} {\bibinfo {author} {\bibfnamefont {J.~K.}\ \bibnamefont
  {Bloomfield}}, \bibinfo {author} {\bibfnamefont {E.~E.}\ \bibnamefont
  {Flanagan}}, \bibinfo {author} {\bibfnamefont {M.}~\bibnamefont {Park}}, \
  and\ \bibinfo {author} {\bibfnamefont {S.}~\bibnamefont {Watson}},\ }\href
  {\doibase 10.1088/1475-7516/2013/08/010} {\bibfield  {journal} {\bibinfo
  {journal} {JCAP}\ }\textbf {\bibinfo {volume} {1308}},\ \bibinfo {pages}
  {010} (\bibinfo {year} {2013})},\ \Eprint {http://arxiv.org/abs/1211.7054}
  {arXiv:1211.7054 [astro-ph.CO]} \BibitemShut {NoStop}%
\bibitem [{\citenamefont {Gleyzes}\ \emph {et~al.}(2013)\citenamefont
  {Gleyzes}, \citenamefont {Langlois}, \citenamefont {Piazza},\ and\
  \citenamefont {Vernizzi}}]{Gleyzes:2013ooa}%
  \BibitemOpen
  \bibfield  {author} {\bibinfo {author} {\bibfnamefont {J.}~\bibnamefont
  {Gleyzes}}, \bibinfo {author} {\bibfnamefont {D.}~\bibnamefont {Langlois}},
  \bibinfo {author} {\bibfnamefont {F.}~\bibnamefont {Piazza}}, \ and\ \bibinfo
  {author} {\bibfnamefont {F.}~\bibnamefont {Vernizzi}},\ }\href {\doibase
  10.1088/1475-7516/2013/08/025} {\bibfield  {journal} {\bibinfo  {journal}
  {JCAP}\ }\textbf {\bibinfo {volume} {1308}},\ \bibinfo {pages} {025}
  (\bibinfo {year} {2013})},\ \Eprint {http://arxiv.org/abs/1304.4840}
  {arXiv:1304.4840 [hep-th]} \BibitemShut {NoStop}%
\bibitem [{\citenamefont {Bloomfield}(2013)}]{Bloomfield:2013efa}%
  \BibitemOpen
  \bibfield  {author} {\bibinfo {author} {\bibfnamefont {J.}~\bibnamefont
  {Bloomfield}},\ }\href {\doibase 10.1088/1475-7516/2013/12/044} {\bibfield
  {journal} {\bibinfo  {journal} {JCAP}\ }\textbf {\bibinfo {volume} {1312}},\
  \bibinfo {pages} {044} (\bibinfo {year} {2013})},\ \Eprint
  {http://arxiv.org/abs/1304.6712} {arXiv:1304.6712 [astro-ph.CO]} \BibitemShut
  {NoStop}%
\bibitem [{\citenamefont {Piazza}\ and\ \citenamefont
  {Vernizzi}(2013)}]{Piazza:2013coa}%
  \BibitemOpen
  \bibfield  {author} {\bibinfo {author} {\bibfnamefont {F.}~\bibnamefont
  {Piazza}}\ and\ \bibinfo {author} {\bibfnamefont {F.}~\bibnamefont
  {Vernizzi}},\ }\href {\doibase 10.1088/0264-9381/30/21/214007} {\bibfield
  {journal} {\bibinfo  {journal} {Class. Quant. Grav.}\ }\textbf {\bibinfo
  {volume} {30}},\ \bibinfo {pages} {214007} (\bibinfo {year} {2013})},\
  \Eprint {http://arxiv.org/abs/1307.4350} {arXiv:1307.4350 [hep-th]}
  \BibitemShut {NoStop}%
\bibitem [{\citenamefont {Bellini}\ and\ \citenamefont
  {Sawicki}(2014)}]{Bellini:2014fua}%
  \BibitemOpen
  \bibfield  {author} {\bibinfo {author} {\bibfnamefont {E.}~\bibnamefont
  {Bellini}}\ and\ \bibinfo {author} {\bibfnamefont {I.}~\bibnamefont
  {Sawicki}},\ }\href {\doibase 10.1088/1475-7516/2014/07/050} {\bibfield
  {journal} {\bibinfo  {journal} {JCAP}\ }\textbf {\bibinfo {volume} {1407}},\
  \bibinfo {pages} {050} (\bibinfo {year} {2014})},\ \Eprint
  {http://arxiv.org/abs/1404.3713} {arXiv:1404.3713 [astro-ph.CO]} \BibitemShut
  {NoStop}%
\bibitem [{\citenamefont {Abbott}\ \emph
  {et~al.}(2017{\natexlab{a}})\citenamefont {Abbott} \emph
  {et~al.}}]{TheLIGOScientific:2017qsa}%
  \BibitemOpen
  \bibfield  {author} {\bibinfo {author} {\bibfnamefont {B.}~\bibnamefont
  {Abbott}} \emph {et~al.} (\bibinfo {collaboration} {Virgo, LIGO
  Scientific}),\ }\href {\doibase 10.1103/PhysRevLett.119.161101} {\bibfield
  {journal} {\bibinfo  {journal} {Phys. Rev. Lett.}\ }\textbf {\bibinfo
  {volume} {119}},\ \bibinfo {pages} {161101} (\bibinfo {year}
  {2017}{\natexlab{a}})},\ \Eprint {http://arxiv.org/abs/1710.05832}
  {arXiv:1710.05832 [gr-qc]} \BibitemShut {NoStop}%
\bibitem [{\citenamefont {Abbott}\ \emph
  {et~al.}(2017{\natexlab{b}})\citenamefont {Abbott} \emph
  {et~al.}}]{Monitor:2017mdv}%
  \BibitemOpen
  \bibfield  {author} {\bibinfo {author} {\bibfnamefont {B.~P.}\ \bibnamefont
  {Abbott}} \emph {et~al.} (\bibinfo {collaboration} {Virgo, Fermi-GBM,
  INTEGRAL, LIGO Scientific}),\ }\href {\doibase 10.3847/2041-8213/aa920c}
  {\bibfield  {journal} {\bibinfo  {journal} {Astrophys. J.}\ }\textbf
  {\bibinfo {volume} {848}},\ \bibinfo {pages} {L13} (\bibinfo {year}
  {2017}{\natexlab{b}})},\ \Eprint {http://arxiv.org/abs/1710.05834}
  {arXiv:1710.05834 [astro-ph.HE]} \BibitemShut {NoStop}%
\bibitem [{\citenamefont {Coulter}\ \emph {et~al.}(2017)\citenamefont {Coulter}
  \emph {et~al.}}]{Coulter:2017wya}%
  \BibitemOpen
  \bibfield  {author} {\bibinfo {author} {\bibfnamefont {D.~A.}\ \bibnamefont
  {Coulter}} \emph {et~al.},\ }\href {\doibase 10.1126/science.aap9811}
  {\bibfield  {journal} {\bibinfo  {journal} {Science}\ } (\bibinfo {year}
  {2017}),\ 10.1126/science.aap9811},\ \Eprint
  {http://arxiv.org/abs/1710.05452} {arXiv:1710.05452 [astro-ph.HE]}
  \BibitemShut {NoStop}%
\bibitem [{\citenamefont {de~Rham}\ and\ \citenamefont
  {Melville}(2018)}]{deRham:2018red}%
  \BibitemOpen
  \bibfield  {author} {\bibinfo {author} {\bibfnamefont {C.}~\bibnamefont
  {de~Rham}}\ and\ \bibinfo {author} {\bibfnamefont {S.}~\bibnamefont
  {Melville}},\ }\href@noop {} {\  (\bibinfo {year} {2018})},\ \Eprint
  {http://arxiv.org/abs/1806.09417} {arXiv:1806.09417 [hep-th]} \BibitemShut
  {NoStop}%
\bibitem [{\citenamefont {Silvestri}\ \emph {et~al.}(2013)\citenamefont
  {Silvestri}, \citenamefont {Pogosian},\ and\ \citenamefont
  {Buniy}}]{Silvestri:2013ne}%
  \BibitemOpen
  \bibfield  {author} {\bibinfo {author} {\bibfnamefont {A.}~\bibnamefont
  {Silvestri}}, \bibinfo {author} {\bibfnamefont {L.}~\bibnamefont {Pogosian}},
  \ and\ \bibinfo {author} {\bibfnamefont {R.~V.}\ \bibnamefont {Buniy}},\
  }\href {\doibase 10.1103/PhysRevD.87.104015} {\bibfield  {journal} {\bibinfo
  {journal} {Phys. Rev.}\ }\textbf {\bibinfo {volume} {D87}},\ \bibinfo {pages}
  {104015} (\bibinfo {year} {2013})},\ \Eprint {http://arxiv.org/abs/1302.1193}
  {arXiv:1302.1193 [astro-ph.CO]} \BibitemShut {NoStop}%
\bibitem [{\citenamefont {Pogosian}\ and\ \citenamefont
  {Silvestri}(2016)}]{Pogosian:2016pwr}%
  \BibitemOpen
  \bibfield  {author} {\bibinfo {author} {\bibfnamefont {L.}~\bibnamefont
  {Pogosian}}\ and\ \bibinfo {author} {\bibfnamefont {A.}~\bibnamefont
  {Silvestri}},\ }\href {\doibase 10.1103/PhysRevD.94.104014} {\bibfield
  {journal} {\bibinfo  {journal} {Phys. Rev.}\ }\textbf {\bibinfo {volume}
  {D94}},\ \bibinfo {pages} {104014} (\bibinfo {year} {2016})},\ \Eprint
  {http://arxiv.org/abs/1606.05339} {arXiv:1606.05339 [astro-ph.CO]}
  \BibitemShut {NoStop}%
\bibitem [{\citenamefont {Kreisch}\ and\ \citenamefont
  {Komatsu}(2017)}]{Kreisch:2017uet}%
  \BibitemOpen
  \bibfield  {author} {\bibinfo {author} {\bibfnamefont {C.~D.}\ \bibnamefont
  {Kreisch}}\ and\ \bibinfo {author} {\bibfnamefont {E.}~\bibnamefont
  {Komatsu}},\ }\href@noop {} {\  (\bibinfo {year} {2017})},\ \Eprint
  {http://arxiv.org/abs/1712.02710} {arXiv:1712.02710 [astro-ph.CO]}
  \BibitemShut {NoStop}%
\bibitem [{\citenamefont {Brans}\ and\ \citenamefont
  {Dicke}(1961)}]{Brans:1961sx}%
  \BibitemOpen
  \bibfield  {author} {\bibinfo {author} {\bibfnamefont {C.}~\bibnamefont
  {Brans}}\ and\ \bibinfo {author} {\bibfnamefont {R.~H.}\ \bibnamefont
  {Dicke}},\ }\href {\doibase 10.1103/PhysRev.124.925} {\bibfield  {journal}
  {\bibinfo  {journal} {Phys. Rev.}\ }\textbf {\bibinfo {volume} {124}},\
  \bibinfo {pages} {925} (\bibinfo {year} {1961})}\BibitemShut {NoStop}%
\bibitem [{\citenamefont {Hu}\ and\ \citenamefont {Sawicki}(2007)}]{Hu:2007nk}%
  \BibitemOpen
  \bibfield  {author} {\bibinfo {author} {\bibfnamefont {W.}~\bibnamefont
  {Hu}}\ and\ \bibinfo {author} {\bibfnamefont {I.}~\bibnamefont {Sawicki}},\
  }\href {\doibase 10.1103/PhysRevD.76.064004} {\bibfield  {journal} {\bibinfo
  {journal} {Phys. Rev.}\ }\textbf {\bibinfo {volume} {D76}},\ \bibinfo {pages}
  {064004} (\bibinfo {year} {2007})},\ \Eprint {http://arxiv.org/abs/0705.1158}
  {arXiv:0705.1158 [astro-ph]} \BibitemShut {NoStop}%
\bibitem [{\citenamefont {Deffayet}\ \emph {et~al.}(2010)\citenamefont
  {Deffayet}, \citenamefont {Pujolas}, \citenamefont {Sawicki},\ and\
  \citenamefont {Vikman}}]{Deffayet:2010qz}%
  \BibitemOpen
  \bibfield  {author} {\bibinfo {author} {\bibfnamefont {C.}~\bibnamefont
  {Deffayet}}, \bibinfo {author} {\bibfnamefont {O.}~\bibnamefont {Pujolas}},
  \bibinfo {author} {\bibfnamefont {I.}~\bibnamefont {Sawicki}}, \ and\
  \bibinfo {author} {\bibfnamefont {A.}~\bibnamefont {Vikman}},\ }\href
  {\doibase 10.1088/1475-7516/2010/10/026} {\bibfield  {journal} {\bibinfo
  {journal} {JCAP}\ }\textbf {\bibinfo {volume} {1010}},\ \bibinfo {pages}
  {026} (\bibinfo {year} {2010})},\ \Eprint {http://arxiv.org/abs/1008.0048}
  {arXiv:1008.0048 [hep-th]} \BibitemShut {NoStop}%
\bibitem [{\citenamefont {Blas}\ \emph {et~al.}(2016)\citenamefont {Blas},
  \citenamefont {Ivanov}, \citenamefont {Sawicki},\ and\ \citenamefont
  {Sibiryakov}}]{Blas:2016qmn}%
  \BibitemOpen
  \bibfield  {author} {\bibinfo {author} {\bibfnamefont {D.}~\bibnamefont
  {Blas}}, \bibinfo {author} {\bibfnamefont {M.~M.}\ \bibnamefont {Ivanov}},
  \bibinfo {author} {\bibfnamefont {I.}~\bibnamefont {Sawicki}}, \ and\
  \bibinfo {author} {\bibfnamefont {S.}~\bibnamefont {Sibiryakov}},\
  }\href@noop {} {\  (\bibinfo {year} {2016})},\ \Eprint
  {http://arxiv.org/abs/1602.04188} {arXiv:1602.04188 [gr-qc]} \BibitemShut
  {NoStop}%
\bibitem [{\citenamefont {Abbott}\ \emph {et~al.}(2016)\citenamefont {Abbott}
  \emph {et~al.}}]{Abbott:2016blz}%
  \BibitemOpen
  \bibfield  {author} {\bibinfo {author} {\bibfnamefont {B.~P.}\ \bibnamefont
  {Abbott}} \emph {et~al.} (\bibinfo {collaboration} {Virgo, LIGO
  Scientific}),\ }\href {\doibase 10.1103/PhysRevLett.116.061102} {\bibfield
  {journal} {\bibinfo  {journal} {Phys. Rev. Lett.}\ }\textbf {\bibinfo
  {volume} {116}},\ \bibinfo {pages} {061102} (\bibinfo {year} {2016})},\
  \Eprint {http://arxiv.org/abs/1602.03837} {arXiv:1602.03837 [gr-qc]}
  \BibitemShut {NoStop}%
\bibitem [{\citenamefont {Hu}\ \emph {et~al.}(2014)\citenamefont {Hu},
  \citenamefont {Raveri}, \citenamefont {Frusciante},\ and\ \citenamefont
  {Silvestri}}]{Hu:2013twa}%
  \BibitemOpen
  \bibfield  {author} {\bibinfo {author} {\bibfnamefont {B.}~\bibnamefont
  {Hu}}, \bibinfo {author} {\bibfnamefont {M.}~\bibnamefont {Raveri}}, \bibinfo
  {author} {\bibfnamefont {N.}~\bibnamefont {Frusciante}}, \ and\ \bibinfo
  {author} {\bibfnamefont {A.}~\bibnamefont {Silvestri}},\ }\href {\doibase
  10.1103/PhysRevD.89.103530} {\bibfield  {journal} {\bibinfo  {journal} {Phys.
  Rev.}\ }\textbf {\bibinfo {volume} {D89}},\ \bibinfo {pages} {103530}
  (\bibinfo {year} {2014})},\ \Eprint {http://arxiv.org/abs/1312.5742}
  {arXiv:1312.5742 [astro-ph.CO]} \BibitemShut {NoStop}%
\bibitem [{\citenamefont {Raveri}\ \emph {et~al.}(2014)\citenamefont {Raveri},
  \citenamefont {Hu}, \citenamefont {Frusciante},\ and\ \citenamefont
  {Silvestri}}]{Raveri:2014cka}%
  \BibitemOpen
  \bibfield  {author} {\bibinfo {author} {\bibfnamefont {M.}~\bibnamefont
  {Raveri}}, \bibinfo {author} {\bibfnamefont {B.}~\bibnamefont {Hu}}, \bibinfo
  {author} {\bibfnamefont {N.}~\bibnamefont {Frusciante}}, \ and\ \bibinfo
  {author} {\bibfnamefont {A.}~\bibnamefont {Silvestri}},\ }\href {\doibase
  10.1103/PhysRevD.90.043513} {\bibfield  {journal} {\bibinfo  {journal} {Phys.
  Rev.}\ }\textbf {\bibinfo {volume} {D90}},\ \bibinfo {pages} {043513}
  (\bibinfo {year} {2014})},\ \Eprint {http://arxiv.org/abs/1405.1022}
  {arXiv:1405.1022 [astro-ph.CO]} \BibitemShut {NoStop}%
\bibitem [{\citenamefont {Lewis}\ \emph {et~al.}(2000)\citenamefont {Lewis},
  \citenamefont {Challinor},\ and\ \citenamefont {Lasenby}}]{Lewis:1999bs}%
  \BibitemOpen
  \bibfield  {author} {\bibinfo {author} {\bibfnamefont {A.}~\bibnamefont
  {Lewis}}, \bibinfo {author} {\bibfnamefont {A.}~\bibnamefont {Challinor}}, \
  and\ \bibinfo {author} {\bibfnamefont {A.}~\bibnamefont {Lasenby}},\
  }\href@noop {} {\bibfield  {journal} {\bibinfo  {journal} {Astrophys. J.}\
  }\textbf {\bibinfo {volume} {538}},\ \bibinfo {pages} {473} (\bibinfo {year}
  {2000})},\ \Eprint {http://arxiv.org/abs/astro-ph/9911177} {astro-ph/9911177}
  \BibitemShut {NoStop}%
\bibitem [{\citenamefont {{Hu}}\ \emph {et~al.}(2014)\citenamefont {{Hu}},
  \citenamefont {{Raveri}}, \citenamefont {{Frusciante}},\ and\ \citenamefont
  {{Silvestri}}}]{2014arXiv1405.3590H}%
  \BibitemOpen
  \bibfield  {author} {\bibinfo {author} {\bibfnamefont {B.}~\bibnamefont
  {{Hu}}}, \bibinfo {author} {\bibfnamefont {M.}~\bibnamefont {{Raveri}}},
  \bibinfo {author} {\bibfnamefont {N.}~\bibnamefont {{Frusciante}}}, \ and\
  \bibinfo {author} {\bibfnamefont {A.}~\bibnamefont {{Silvestri}}},\
  }\href@noop {} {\bibfield  {journal} {\bibinfo  {journal} {ArXiv e-prints}\ }
  (\bibinfo {year} {2014})},\ \Eprint {http://arxiv.org/abs/1405.3590}
  {arXiv:1405.3590 [astro-ph.IM]} \BibitemShut {NoStop}%
\bibitem [{\citenamefont {Uzan}(2011)}]{Uzan:2010pm}%
  \BibitemOpen
  \bibfield  {author} {\bibinfo {author} {\bibfnamefont {J.-P.}\ \bibnamefont
  {Uzan}},\ }\href {\doibase 10.12942/lrr-2011-2} {\bibfield  {journal}
  {\bibinfo  {journal} {Living Rev. Rel.}\ }\textbf {\bibinfo {volume} {14}},\
  \bibinfo {pages} {2} (\bibinfo {year} {2011})},\ \Eprint
  {http://arxiv.org/abs/1009.5514} {arXiv:1009.5514 [astro-ph.CO]} \BibitemShut
  {NoStop}%
\bibitem [{\citenamefont {Brax}\ \emph {et~al.}(2012)\citenamefont {Brax},
  \citenamefont {Davis}, \citenamefont {Li},\ and\ \citenamefont
  {Winther}}]{Brax:2012gr}%
  \BibitemOpen
  \bibfield  {author} {\bibinfo {author} {\bibfnamefont {P.}~\bibnamefont
  {Brax}}, \bibinfo {author} {\bibfnamefont {A.-C.}\ \bibnamefont {Davis}},
  \bibinfo {author} {\bibfnamefont {B.}~\bibnamefont {Li}}, \ and\ \bibinfo
  {author} {\bibfnamefont {H.~A.}\ \bibnamefont {Winther}},\ }\href {\doibase
  10.1103/PhysRevD.86.044015} {\bibfield  {journal} {\bibinfo  {journal} {Phys.
  Rev.}\ }\textbf {\bibinfo {volume} {D86}},\ \bibinfo {pages} {044015}
  (\bibinfo {year} {2012})},\ \Eprint {http://arxiv.org/abs/1203.4812}
  {arXiv:1203.4812 [astro-ph.CO]} \BibitemShut {NoStop}%
\bibitem [{\citenamefont {Joyce}\ \emph {et~al.}(2015)\citenamefont {Joyce},
  \citenamefont {Jain}, \citenamefont {Khoury},\ and\ \citenamefont
  {Trodden}}]{Joyce:2014kja}%
  \BibitemOpen
  \bibfield  {author} {\bibinfo {author} {\bibfnamefont {A.}~\bibnamefont
  {Joyce}}, \bibinfo {author} {\bibfnamefont {B.}~\bibnamefont {Jain}},
  \bibinfo {author} {\bibfnamefont {J.}~\bibnamefont {Khoury}}, \ and\ \bibinfo
  {author} {\bibfnamefont {M.}~\bibnamefont {Trodden}},\ }\href {\doibase
  10.1016/j.physrep.2014.12.002} {\bibfield  {journal} {\bibinfo  {journal}
  {Phys. Rept.}\ }\textbf {\bibinfo {volume} {568}},\ \bibinfo {pages} {1}
  (\bibinfo {year} {2015})},\ \Eprint {http://arxiv.org/abs/1407.0059}
  {arXiv:1407.0059 [astro-ph.CO]} \BibitemShut {NoStop}%
\bibitem [{\citenamefont {Perenon}\ \emph {et~al.}(2015)\citenamefont
  {Perenon}, \citenamefont {Piazza}, \citenamefont {Marinoni},\ and\
  \citenamefont {Hui}}]{Perenon:2015sla}%
  \BibitemOpen
  \bibfield  {author} {\bibinfo {author} {\bibfnamefont {L.}~\bibnamefont
  {Perenon}}, \bibinfo {author} {\bibfnamefont {F.}~\bibnamefont {Piazza}},
  \bibinfo {author} {\bibfnamefont {C.}~\bibnamefont {Marinoni}}, \ and\
  \bibinfo {author} {\bibfnamefont {L.}~\bibnamefont {Hui}},\ }\href {\doibase
  10.1088/1475-7516/2015/11/029} {\bibfield  {journal} {\bibinfo  {journal}
  {JCAP}\ }\textbf {\bibinfo {volume} {1511}},\ \bibinfo {pages} {029}
  (\bibinfo {year} {2015})},\ \Eprint {http://arxiv.org/abs/1506.03047}
  {arXiv:1506.03047 [astro-ph.CO]} \BibitemShut {NoStop}%
\bibitem [{\citenamefont {Ade}\ \emph {et~al.}(2016)\citenamefont {Ade} \emph
  {et~al.}}]{Ade:2015xua}%
  \BibitemOpen
  \bibfield  {author} {\bibinfo {author} {\bibfnamefont {P.~A.~R.}\
  \bibnamefont {Ade}} \emph {et~al.} (\bibinfo {collaboration} {Planck}),\
  }\href {\doibase 10.1051/0004-6361/201525830} {\bibfield  {journal} {\bibinfo
   {journal} {Astron. Astrophys.}\ }\textbf {\bibinfo {volume} {594}},\
  \bibinfo {pages} {A13} (\bibinfo {year} {2016})},\ \Eprint
  {http://arxiv.org/abs/1502.01589} {arXiv:1502.01589 [astro-ph.CO]}
  \BibitemShut {NoStop}%
\bibitem [{\citenamefont {{Suzuki}}(2012)}]{2012ApJ...746...85S}%
  \BibitemOpen
  \bibfield  {author} {\bibinfo {author} {\bibfnamefont {T.}~\bibnamefont
  {{Suzuki}}, \bibfnamefont {N.~et~al}},\ }\href {\doibase
  10.1088/0004-637X/746/1/85} {\bibfield  {journal} {\bibinfo  {journal}
  {\apj}\ }\textbf {\bibinfo {volume} {746}},\ \bibinfo {eid} {85} (\bibinfo
  {year} {2012})},\ \Eprint {http://arxiv.org/abs/1105.3470} {arXiv:1105.3470
  [astro-ph.CO]} \BibitemShut {NoStop}%
\bibitem [{\citenamefont {Lewis}\ and\ \citenamefont
  {Bridle}(2002)}]{Lewis:2002ah}%
  \BibitemOpen
  \bibfield  {author} {\bibinfo {author} {\bibfnamefont {A.}~\bibnamefont
  {Lewis}}\ and\ \bibinfo {author} {\bibfnamefont {S.}~\bibnamefont {Bridle}},\
  }\href@noop {} {\bibfield  {journal} {\bibinfo  {journal} {Phys. Rev.}\
  }\textbf {\bibinfo {volume} {D66}},\ \bibinfo {pages} {103511} (\bibinfo
  {year} {2002})},\ \Eprint {http://arxiv.org/abs/astro-ph/0205436}
  {astro-ph/0205436} \BibitemShut {NoStop}%
\end{thebibliography}%

\end{document}